%% file: main.tex
\documentclass[11pt]{article}

\usepackage{a4}
\usepackage{authblk}
\usepackage{amsmath,amssymb}
\usepackage{amsthm}
\usepackage[dvipdfmx]{graphicx}

\usepackage{comment}
\usepackage{ascmac}
\usepackage{tabls}
\usepackage{booktabs}
\usepackage[dvipdfmx]{graphicx}
\usepackage{wrapfig}
\usepackage{array}
\usepackage[algoruled,linesnumbered,algo2e,vlined]{algorithm2e}
\usepackage{url}
\usepackage{multirow}
\usepackage{multicol}
\usepackage{cases}
\usepackage[normalem]{ulem}
\usepackage{boites}

\usepackage{here}

\usepackage[pagewise,mathlines]{lineno}

\usepackage[final,mode=multiuser]{fixme}

\usepackage{colortbl}

\fxsetup{theme=color}
\definecolor{fxtarget}{rgb}{0.0000,0.0000,0.4823}

\fxuseenvlayout{colorsig}
\fxusetargetlayout{color}

\fxuseenvlayout{colorsig}
\fxusetargetlayout{color}
\FXRegisterAuthor{si}{asi}{SI}
\fxsetface{env}{}

\pdfoutput=1

\usepackage{thmtools}

\newtheorem{theorem}{Theorem}
\newtheorem{lemma}{Lemma}
\newtheorem{corollary}{Corollary}

\theoremstyle{definition}

\newtheorem{remark}{Remark}
\newtheorem{example}{Example}

\newcommand{\Prefix}{\mathsf{Prefix}}
\newcommand{\Substr}{\mathsf{Substr}}
\newcommand{\Suffix}{\mathsf{Suffix}}

\newcommand{\rev}[1]{#1^R}

\newcommand{\LeftM}{\mathsf{LeftM}}
\newcommand{\RightM}{\mathsf{RightM}}
\newcommand{\M}{\mathsf{M}}

\newcommand{\STrie}{\mathsf{STrie}}
\newcommand{\STree}{\mathsf{STree}}
\newcommand{\CDAWG}{\mathsf{CDAWG}}

\newcommand{\LST}{\mathsf{LSTrie}}
\newcommand{\simLST}{\mathsf{simLSTrie}}
\newcommand{\LCDAWG}{\mathsf{LCDAWG}}
\newcommand{\simLCDAWG}{\mathsf{simLCDAWG}}

\newcommand{\VSTrie}{\mathsf{V_{\STrie}}}
\newcommand{\ESTrie}{\mathsf{E_{\STrie}}}
\newcommand{\VSTree}{\mathsf{V_{\STree}}}
\newcommand{\ESTree}{\mathsf{E_{\STree}}}
\newcommand{\VCDAWG}{\mathsf{V_{\CDAWG}}}
\newcommand{\ECDAWG}{\mathsf{E_{\CDAWG}}}

\newcommand{\VLST}{\mathsf{V_{\LST}}}
\newcommand{\VLSTone}{\mathsf{V^1_{\LST}}}
\newcommand{\VLSTtwo}{\mathsf{V^2_{\LST}}}
\newcommand{\ELST}{\mathsf{E_{\LST}}}

\newcommand{\VsimLST}{\mathsf{V_{\simLST}}}
\newcommand{\EsimLST}{\mathsf{E_{\simLST}}}

\newcommand{\VLCDAWG}{\mathsf{V_{\LCDAWG}}}
\newcommand{\VLCDAWGone}{\mathsf{V^1_{\LCDAWG}}}
\newcommand{\VLCDAWGtwo}{\mathsf{V^2_{\LCDAWG}}}
\newcommand{\ELCDAWG}{\mathsf{E_{\LCDAWG}}}

\newcommand{\VsimLCDAWG}{\mathsf{V_{\simLCDAWG}}}
\newcommand{\EsimLCDAWG}{\mathsf{E_{\simLCDAWG}}}

\newcommand{\righte}{\mathsf{er}}
\newcommand{\lefte}{\mathsf{el}}

\newcommand{\str}{\mathsf{str}}
\newcommand{\Plus}{\mathsf{+}}
\newcommand{\slink}{\mathsf{slink}}
\newcommand{\fastlink}{\mathsf{flink}}

\newcommand{\LPT}{\mathsf{LPT}}
\newcommand{\LPTplus}{\mathsf{LPT}^+}

\newcommand{\heightSLT}{\mathsf{h}_{\slink}}

\newcommand{\occ}{\mathsf{occ}}

\begin{document}

\title{Linear-size Suffix Tries and Linear-size CDAWGs \\ Simplified and Improved}

\author{Shunsuke Inenaga}

\affil{Department of Informatics, Kyushu University}

\date{}
\maketitle

\begin{abstract}
  The \emph{linear-size suffix tries} (\emph{LSTries}) [Crochemore et al., TCS 2016] are a version of suffix trees in which the edge labels are single characters, yet are able to perform pattern matching queries in optimal time. Instead of explicitly storing the input text, LSTries have some extra non-branching internal nodes called \emph{type-2} nodes. The extended techniques are then used in the \emph{linear-size compact directed acyclic word graphs} (\emph{LCDAWGs}) [Takagi et al. SPIRE 2017], which can be stored with $O(\lefte(T)+\righte(T))$ space (i.e. without the text), where $\lefte(T)$ and $\righte(T)$ are the numbers of left- and right-extensions of the maximal repeats in the input text string $T$, respectively.
  In this paper, we present simpler alternatives to the aforementioned indexing structures, called the \emph{simplified LSTries} (\emph{simLSTries}) and the \emph{simplified LCDAWGs} (\emph{simLCDAWGs}), in which most of the type-2 nodes are removed. In particular, our simLCDAWGs require only $O(\righte(T))$ space and work in a weaker model of computation (i.e. the pointer machine model). This contrasts the $O(\righte(T))$-space CDAWG representation of [Belazzougui \& Cunial, SPIRE 2017], which works on the word RAM model. 
\end{abstract}

\input{introduction.tex}
\input{preliminaries.tex}
\input{lst.tex}
\input{sim_lst.tex}
\input{lcdawg.tex}
\input{sim_lcdawg.tex}
\input{conclusions.tex}

\section*{Acknowledgments}
This work was supported by JSPS KAKENHI Grant Numbers JP23K24808 and JP23K18466.
The author thanks the anonymous referees for their helpful comments and suggestions.

\bibliographystyle{abbrv}
\bibliography{ref}

\clearpage

\end{document}

%% file: introduction.tex
\section{Introduction}

Text indexing is a central task in string processing with applications to a variety of fields including information retrieval, bioinformatics, data compression, and data mining.
The goal is to preprocess the given text string $T$ so that later, given a query pattern $P$, one can quickly find the occurrences of $P$ in $T$.

The fundamental text indexing structure is the \emph{suffix tree}~\cite{Weiner1973}, which is an edge-labeled rooted tree that represents all suffixes of the input text $T$. The \emph{compact directed acyclic word graph} (\emph{CDAWG})~\cite{Blumer1987} is a space-efficient alternative, which can be obtained by merging isomorphic subtrees of the suffix tree.
Both the suffix tree and the CDAWG for $T$ support pattern matching queries in optimal $O(m \log \sigma + \occ)$ time for a query pattern $P$ of length $m$, where $\sigma$ denotes the alphabet size and $\occ$ denotes the number of occurrences of $P$ in $T$.
While the number of nodes and edges of suffix trees is $\Theta(n)$ for any string $T$ of length $n$,
CDAWGs may have much less nodes and edges for highly repetitive strings~\cite{Rytter06,RadoszewskiR12} since there tend to exist many and large isomorphic subtrees in their suffix trees.
Thus the number $\righte(T)$ of CDAWG edges serves as one of the string repetitiveness measures~\cite{Navarro21a}.
Although $\righte(T)$ itself is a weaker measure than the well-studied other measures including the Lempel-Ziv 77 parse~\cite{LZ77}, the (smallest) grammar compressors~\cite{Rytter03,CharikarLLPPSS05}, and the string attractors~\cite{KempaP18}, CDAWGs are suited for pattern matching due to the connection to suffix trees~\cite{Navarro21a}.
Also, the equivalence classes of substrings induced from the CDAWG nodes are known to be useful in string pattern discovery~\cite{TakedaFNYT03}.

Classical representations of suffix trees and CDAWGs require to store the input text $T$ explicitly, since the string label $x$ of each edge is represented by a pair $(i,j)$ of positions in $T$ such that the substring $T[i..j]$ is equal to $x$. This enables us to represent the suffix tree in $O(n)$ space, and the CDAWG in $O(n + \righte(T)) \subseteq O(n)$ space.

The \emph{linear-size suffix trie} (\emph{LSTrie})~\cite{Crochemore2016} is a variant of the suffix-tree based indexing structure that does not require to explicitly store the input string $T$, yet supporting (the decision version of) pattern matching queries in $O(m \log \sigma)$ time.
The idea of LSTries was to enhance suffix trees with some extra non-branching nodes called \emph{type-2 nodes}, so that the edge labels can be retrieved by navigating suffix links of edges (called \emph{fast links}).
It is known that the total number of nodes in LSTries is at most $3n-3$ (Lemma~\ref{lem:LST_size}).

The \emph{linear-size CDAWG} (\emph{LCDAWG})~\cite{Takagi2017} is the CDAWG version of the same kind (i.e. using type-2 nodes), which supports $O(m \log \sigma + \occ)$-time pattern matching and occupies $O(\lefte(T)+\righte(T))$ space, where $\lefte(T)$ is the number of edges in the CDAWG of the reversal of $T$.

In this paper, we revisit LSTries and LCDAWGs and show that (most) type-2 nodes are not necessary to support pattern matching queries, even without the text.
Our proposed indexing structures, called the \emph{simplified LSTrie} (\emph{simLSTrie}) and the \emph{simplified LCDAWG} (\emph{simLCDAWG}), both support (the reporting version of) pattern matching queries in optimal $O(m \log \sigma + \occ)$ time.

\begin{table}[t]
  \caption{Comparison of the numbers of nodes and edges in the LSTrie-based data structures. While our simLSTries can report all $\occ$ occurrences of a given pattern $P$ in $O(\occ)$ time after locating the locus in $O(m \log \sigma)$ time, the previous LSTries need extra $n-1$ edges (and hence at most $4n-5$ edges) to report occurrences in $O(\occ)$ time (c.f. Remark~\ref{rem:lst_occ}). The $\log \sigma$ factor, where $\sigma$ is the alphabet size, can be removed if one uses hashing.}
  \label{tbl:comparisons_LST}
  \centering
  \begin{tabular}{c||l|c|c}
    \hline
    data structure & \# nodes & \# edges & model of computation \\
    \hline \hline
    LSTries~\cite{Crochemore2016} & $\leq 3n-3$ & $\leq 3n-4$ & pointer machine \\
    \hline
    simLSTrie~[this work] & $\leq 2n$ & $\leq 2n-1$ & pointer machine \\
    \hline
  \end{tabular}
\end{table}

As for their size bounds, we show that the number of nodes in simLSTries is at most $2n$ (Lemma~\ref{lem:simLST_size}). See also Table~\ref{tbl:comparisons_LST}.
We then show that simLCDAWGs can be stored with $O(\righte(T))$ space, thus improving on the $O(\lefte(T) + \righte(T))$ space requirement by the original LCDAWGs.
This improvement is supported by the existence of a family of strings for which $\lefte(T) = \Omega(\righte(T)\sqrt{n})$ (Lemma~\ref{lem:lowerbound_el_er}).
Both of our proposed indexing structures are simple enough to work
in the aforementioned complexities even on a weaker model of computation (i.e. the pointer machine model).
We note that our data structures require no complicated manipulations
on the pointer machine model (and thus as well as on the word RAM model).

Belazzougui and Cunial~\cite{BelazzouguiC17} proposed a different $O(\righte(T))$-space representation of the CDAWG that supports pattern matching in $O(m \log \sigma + \occ)$ time\footnote{The original version of Belazzougui and Cunial~\cite{BelazzouguiC17} uses hashing to achieve $O(m + \occ)$-time pattern matching. The same can be done with every indexing structure mentioned here.}.
They represent CDAWG edge labels with a grammar of size $O(\righte(T))$ that is induced from the reversed DAG of the CDAWG.
Their structure needs to access the characters of the decompressed strings in a sequential manner (from left to right) for pattern matching and edge-label extractions.
For this purpose, they enhance this grammar with the constant-time level ancestor (LA)  data structure~\cite{BerkmanV94,BenderF04} that works in the word RAM model.
LCDAWGs~\cite{Takagi2017} also use grammars for representing edge labels, which are enhanced with Gasieniec et al.'s algorithm~\cite{GasieniecKPS05} that is based on the constant-time lowest common ancestor (LCA) data structure~\cite{BenderF00} on the word RAM model.
See Table~\ref{tbl:comparisons_LCDAWG} for a summary of these comparisons.

In addition, we present how to efficiently construct our simLCDAWG from the CDAWG and/or the string (Lemma~\ref{lem:simLCDAWG_construction_from_CDAWG} and Corollary~\ref{coro:simLCDAWG_construction_from_string} in Section~\ref{sec:sim_LCDAWG}).

\begin{table}[t]
  \caption{Comparison of space usage (i.e. the number of edges) and model of computation for the CDAWG-based data structures, each of which can find all $\occ$ occurrences of a given pattern of length $m$ in the text $T$ of length $n$ in $O(m \log \sigma + \occ)$ time. The $\log \sigma$ factor, where $\sigma$ is the alphabet size, can be removed if one uses hashing.}
  \label{tbl:comparisons_LCDAWG}
  \centering
  \begin{tabular}{c||c|c}
    \hline
    data structure & space & model of computation \\
    \hline \hline
    CDAWG~\cite{Blumer1987} & $O(n)$ & pointer machine \\
    \hline
    LCDAWG~\cite{Takagi2017} & $O(\lefte(T)+\righte(T))$ & word RAM \\
    \hline
    Belazzougui \& Cunial~\cite{BelazzouguiC17} & $O(\righte(T))$ & word RAM \\
    \hline
    simLCDAWG~[this work] & $O(\righte(T))$ & pointer machine \\
    \hline
  \end{tabular}
\end{table}

The rest of this paper is organized as follows.
Section~\ref{sec:preliminaries} gives basic notations on strings
and the definitions of classical indexing structures (suffix tries, suffix trees, and CDAWGs).
Section~\ref{sec:lst} reviews LSTries and describes how type-2 nodes are used.
Section~\ref{sec:sim_LST} introduces our first proposed indexing structure, simLSTries.
Section~\ref{sec:lcdawg} reviews LCDAWGs and describes how type-2 nodes are used.
Section~\ref{sec:sim_LCDAWG} introduces our second proposed indexing structure, simLCDAWGs.
We conclude in Section~\ref{sec:conclusions}.

%% file: preliminaries.tex
\section{Preliminaries} \label{sec:preliminaries}

\subsection{Strings}
Let $\Sigma$ denote an ordered \emph{alphabet} of size $\sigma$.
An element of $\Sigma^*$ is called a \emph{string}.
The length of a string $T \in \Sigma^*$ is denoted by $|T|$.
The \emph{empty string} $\varepsilon$ is the string of length $0$.
For string $T = xyz$, $x$, $y$, and $z$ are called
the \emph{prefix}, \emph{substring}, and \emph{suffix} of $T$,
respectively.
Let $\Prefix(T)$, $\Substr(T)$, and $\Suffix(T)$ denote
the sets of prefixes, substrings, and suffixes of $T$, respectively.
The elements of $\Prefix(T) \setminus \{T\}$,
$\Substr(T) \setminus \{T\}$, and $\Suffix(T) \setminus \{T\}$ are called
the \emph{proper prefixes}, \emph{proper substrings},
and \emph{proper suffixes} of $T$, respectively.
For a string $T$ of length $n$, $T[i]$ denotes the $i$-th symbol of $T$
and $T[i..j] = T[i] \cdots T[j]$ denotes the substring of $T$
that begins at position $i$ and ends at position $j$ for $1 \leq i \leq j \leq n$.
For convenience, let $T[i..j] = \varepsilon$ for $i > j$.
The \emph{reverse image} of a string $T$ is denoted by $\rev{T}$,
that is, $\rev{T} = T[|T|] \cdots T[1]$.


\subsection{Maximal Substrings}

A substring $u$ of a string $T$ is said to be \emph{left-maximal} in $T$ iff
(1) there are distinct characters $a, b$ such that $au, bu \in \Substr(T)$, or
(2) $u \in \Prefix(T)$.
For each left-maximal substring $u$ of a string $T$,
the characters $c \in \Sigma$ such that $cu \in \Substr(T)$
are called \emph{left-extensions} of the left-maximal substring $u$.
Similarly, a substring $u$ of $T$ is said to be \emph{right-maximal} in $T$ iff
(1) there are distinct characters $a, b$ such that $ua, ub \in \Substr(T)$, or
(2) $u \in \Suffix(T)$.
For each right-maximal substring $u$ of a string $T$,
the characters $c \in \Sigma$ such that $uc \in \Substr(T)$
are called \emph{right-extensions} of the right-maximal substring $u$.

A substring $u$ of $T$ is said to be \emph{maximal} in $T$ iff
$u$ is both left-maximal and right-maximal in $T$.
Let $\LeftM(T)$, $\RightM(T)$, $\M(T)$ denote the sets of
left-maximal, right-maximal, and maximal substrings of $T$.
By definition, $\M(T) = \LeftM(T) \cap \RightM(T)$ holds.

\subsection{Tries}

A \emph{trie} $\mathsf{T}$ is a rooted tree $(\mathsf{V}, \mathsf{E})$ such that 
(1) each edge in $\mathsf{E}$ is labeled by a single character from $\Sigma$ and
(2) the character labels of the out-going edges of each node
begin with distinct characters.

For a pair of nodes $u,v$ in a trie $\mathsf{T}$ such that $u$ is an ancestor of $v$,
let $\langle u, v \rangle$ denote the path from $u$ to $v$,
and the \emph{path label} of $\langle u, v \rangle$,
denoted $\str(\langle u, v \rangle)$,
is the concatenation of the edge labels from $u$ to $v$.
The \emph{string label} of a node $v$ in a trie $\mathsf{T}$, denoted $\str(v)$,
is the path label from the root to $v$.
The \emph{string depth} of a node $v$ in $\mathsf{T}$ is the length
of the string label of $v$, that is $|\str(v)|$.

We denote each edge in $\mathsf{E}$ by a tuple $(u, a, v)$,
where $u$ is the origin node, $a$ is the character label,
and $v$ is the destination node of the edge.
We sometimes denote the edge from $u$ to $v$ simply by a pair $(u, v)$ 
when its label is not important in the context.


\subsection{Suffix Tries}

The \emph{suffix trie} $\STrie(T) = (\VSTrie, \ESTrie)$ of a string $T$ is a trie that represents $\Substr(T)$.
That is, each substring $u$ of $T$ is spelled out by a unique path from the root of $\STrie(T)$.
Formally, we have that $\Substr(T) = \{ \str(v) \mid v \in \VSTrie \}$,
meaning that there is a one-to-one correspondence between
the nodes of $\STrie(T)$ and the substrings in $\Substr(T)$.
The \emph{suffix link} of a non-root node $v$ in $\STrie(T)$
is an auxiliary edge that points to the node $u$
such that $\str(u) = \str(v)[2..|\str(v)|]$,
which is denoted by $\slink(v) = u$.
By using a common convention that every string $T$ terminates with a unique end-marker
$\$$ that does not occur elsewhere in $T$,
there is a one-to-one correspondence between the suffixes in $\Suffix(T)$
and the leaves in $\STrie(T)$.

Since $|\Substr(T)| = O(n^2)$ for a string of length $n$,
$\STrie(T)$ can have $O(n^2)$ nodes and edges, which is a major drawback of suffix tries.

\subsection{Suffix Trees}

The \emph{suffix tree}~\cite{Weiner1973} of a string $T$,
denoted $\STree(T) = (\VSTree, \ESTree)$, is a path-compressed trie representing $\Suffix(T)$ such that $\RightM(T) = \{\str(v) \mid v \in \VSTree\}$,
meaning that there is a one-to-one correspondence between
the nodes of $\STree(T)$ and the right-maximal substrings in $\RightM(T)$.
By definition, each internal node has at least two children,
each edge is labeled by a non-empty substring of $T$,
and the out-going edges of the same node begin with distinct characters.
The first characters of the labels of the out-going edges of a non-leaf node $v$
are the right-extensions of the corresponding right-maximal substring $\str(v)$.
For the nodes of suffix trees, we use the same conventions for their path labels, string depths, and suffix links as in the case of suffix tries.

Intuitively, $\STree(T)$ can be obtained by keeping only the branching nodes
of $\STrie(T)$ and compressing unary paths of $\STrie(T)$.
Since $\STree(T)$ has $n$ leaves (assuming the end-marker $\$$)
and since its internal nodes are all branching,
$\STree(T)$ has at most $2n-1$ nodes and $2n-2$ edges.
We can represent $\STree(T)$ in $O(n)$ space
by storing the original input string $T$,
and by representing each edge label $x$ with a tuple $(a, (i,j))$ such that
$a = T[i]$ is the first character of $x$ and $T[i..j] = x$.

\sinote*{added}{%
The nodes of $\STree(T)$ are also called \emph{explicit nodes},
and the nodes of $\STrie(T)$ that do not exist in $\STree(T)$ are
called \emph{implicit nodes}.
The \emph{locus} of a substring $w \in \Substr(T)$
is the (explicit or implicit) node $v$ such that $\str(v) = w$.
When the node $v$ is an implicit node, then the locus for $w$ is on an edge in $\STree(T)$.
}%

\begin{figure}[t]
	\centering
	\begin{minipage}[t]{0.3\hsize}
		\centering
		\includegraphics[scale=0.35]{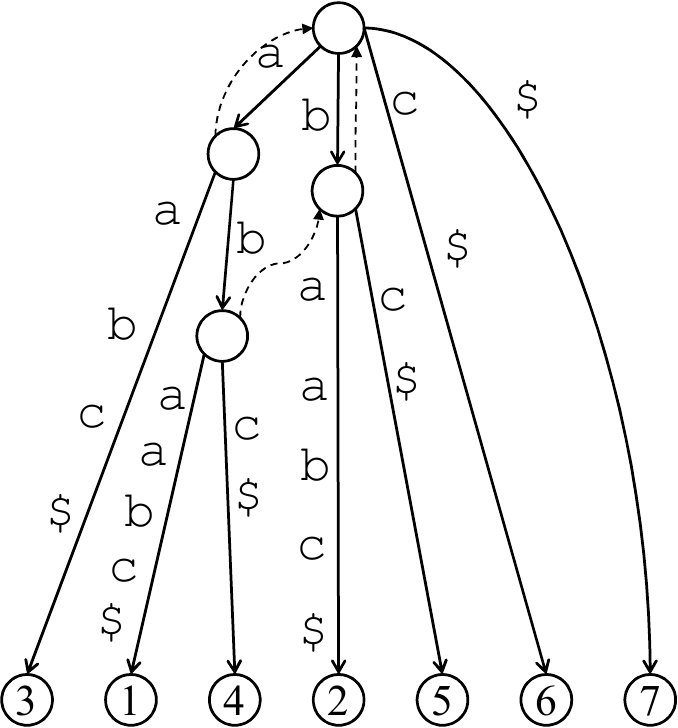}\\
		\ \ \ \small{$\STree(T)$}
	\end{minipage}
        \hfill
	\begin{minipage}[t]{0.3\hsize}
		\centering
		\includegraphics[scale=0.35]{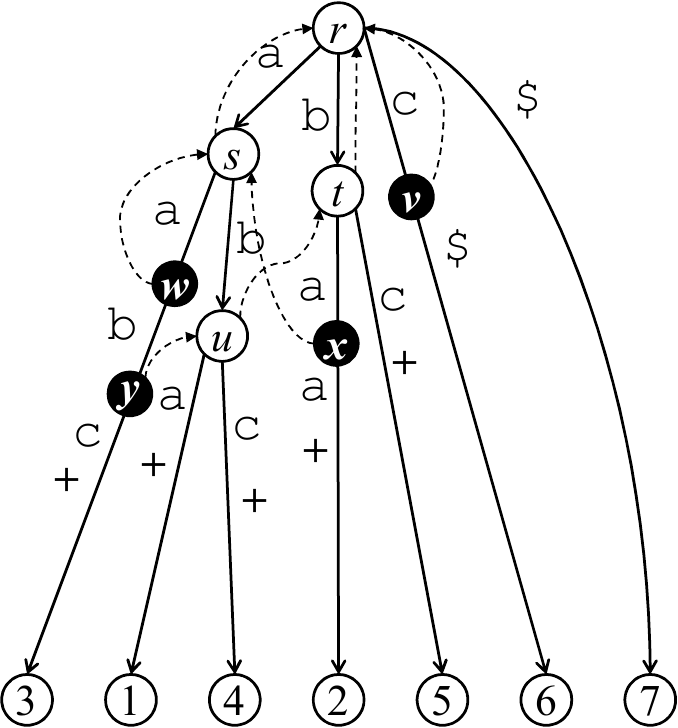}\\
		\ \ \ \small{$\LST(T)$}
	\end{minipage}
        \hfill
	\begin{minipage}[t]{0.3\hsize}
		\centering
		\includegraphics[scale=0.35]{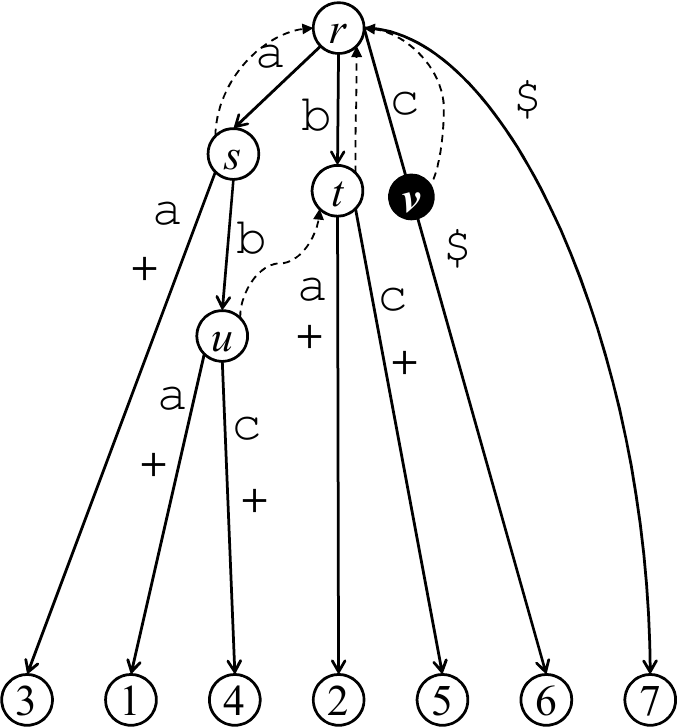}\\
		\ \ \ \small{$\simLST(T)$}
	\end{minipage}
	\caption{
		The suffix tree (STree), linear-size suffix trie (LSTrie), and simplified linear-size suffix trie (simLSTrie) of string $T = \mathtt{abaabc\texttt{\$}}$. The white and black circles represent type-1 and type-2 nodes, respectively. The broken arcs represent suffix links.
	}
	\label{fig:suffix_tree_lstrie}
\end{figure}

See the left diagram of Figure~\ref{fig:suffix_tree_lstrie} for an example of suffix trees.

\subsection{CDAWGs}

The \emph{compact directed acyclic word graph} (\emph{CDAWG}) of a string $T$,
denoted $\CDAWG(T) = (\VCDAWG, \ECDAWG)$,
is a path-compressed smallest partial DFA that represents $\Suffix(T)$,
such that there is a one-to-one correspondence between
the nodes of $\CDAWG(T)$ and the maximal substrings in $\M(T)$.
Namely, for each node $u \in \VCDAWG$ of $\CDAWG(T)$,
$x_u \in \Sigma^*$ is the longest string represented by the node $u$
iff $x_u \in \M(T)$.
The first characters of the labels of the out-going edges of an internal node
of $\CDAWG(T)$ are the right-extensions of the corresponding maximal substring
for that node.
This means that the number $|\ECDAWG|$ of edges in $\CDAWG(T)$
is equal to the total number of right-extensions of maximal substrings $\M(T)$,
which is denoted by $\righte(T)$.
Similarly, we denote by $\lefte(T)$ the number of left-exertions of
the maximal substrings $\M(T)$.
We remark that $\lefte(T) = \righte(\rev{T})$ and $\righte(T) = \lefte(\rev{T})$.

Intuitively, $\CDAWG(T)$ can be obtained by merging isomorphic subtrees of $\STree(T)$, which are connected by a chain of suffix links.
Therefore, each string represented by a node $u$ of $\CDAWG(T)$ is a suffix of all the other longer strings represented by the same node $u$.
For convenience, for any CDAWG node $u \in \VCDAWG$,
let $\str(u)$ denote the \emph{longest} string represented by the node $u$.
The \emph{suffix link} of a CDAWG node $u$ points to another CDAWG node $v$
iff $\str(v)$ is the longest proper suffix of $\str(u)$ that is not represented by $u$.

For each edge $(u, y, v) \in \ECDAWG$ that is labeled by a non-empty string $y$,
the edge $(u, y, v)$ is called a \emph{primary edge}
if $|\str(u)|+|y| = |\str(v)|$,
and $(u, y, v)$ is called a \emph{secondary edge}
otherwise ($|\str(u)|+|y| < |\str(v)|$).
Namely, if we define the weight of each edge as its string label length, 
then the primary edges are the edges in the \emph{longest} paths from the source to all the nodes of $\CDAWG(T)$.
By definition, there are exactly $|\VCDAWG|-1$ primary edges in $\CDAWG(T)$.
Let $\LPT(T)$ denote the spanning tree of $\CDAWG(T)$ that consists of the primary edges.

By representing each edge label with a tuple $(a, (i,j))$ as in the case of suffix trees, and by storing the input string $T$, $\CDAWG(T)$ can be represented in $O(n + \righte(T)) \subseteq O(n)$ space, where the $n$ term is for explicitly storing the input string $T$.

\begin{figure}[t]
	\centering
	\begin{minipage}[t]{0.3\hsize}
		\centering
		\includegraphics[scale=0.35]{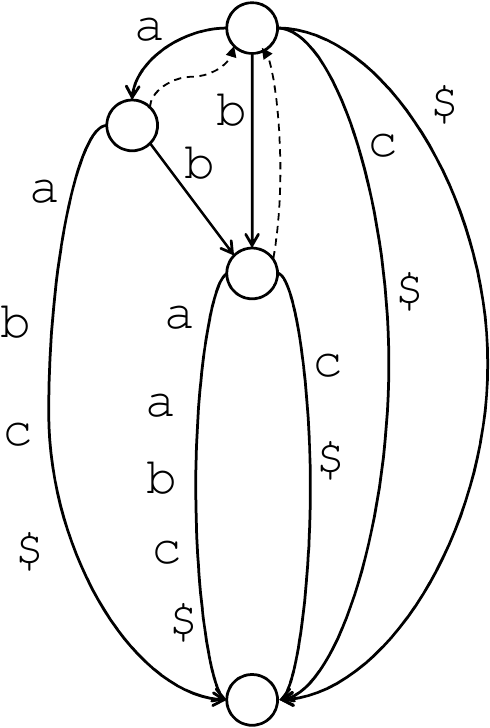}\\
		\ \ \ \small{$\CDAWG(T)$}
	\end{minipage}
        \hfill
	\begin{minipage}[t]{0.3\hsize}
		\centering
		\includegraphics[scale=0.35]{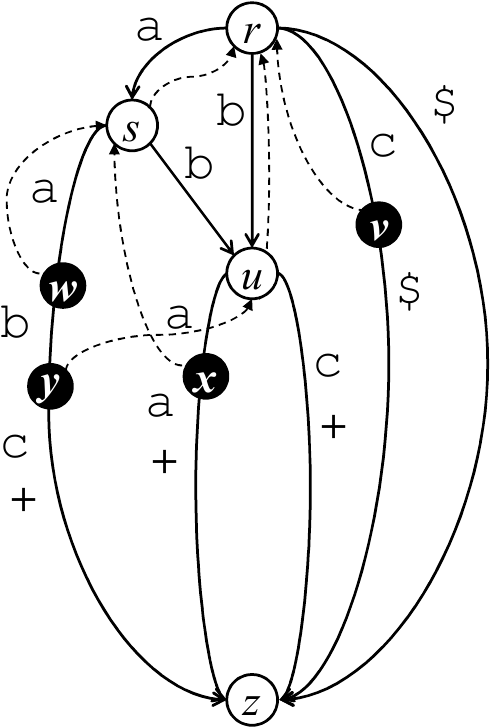}\\
		\ \ \ \small{$\LCDAWG(T)$}
	\end{minipage}
        \hfill
	\begin{minipage}[t]{0.3\hsize}
		\centering
		\includegraphics[scale=0.35]{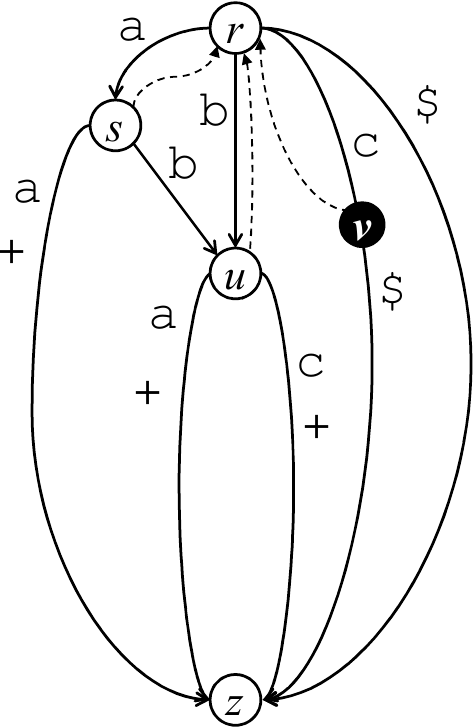}\\
		\ \ \ \small{$\simLCDAWG(T)$}
	\end{minipage}
	\caption{
		The CDAWG, linear-size CDAWG (LCDAWG), and simplified linear-size CDAWG (simLCDAWG) of string $T = \mathtt{abaabc\texttt{\$}}$. The white and black circles represent type-1 and type-2 nodes, respectively. The broken arcs represent suffix links.
	}
	\label{fig:cdawg_lcdawg}
\end{figure}

See the left diagram of Figure~\ref{fig:cdawg_lcdawg} for an example of CDAWGs.

%% file: lst.tex
\section{Linear-size Suffix Tries} \label{sec:lst}

This section reviews 
the \emph{linear-size suffix trie} (\emph{LSTrie}) data structure,
originally proposed by Crochemore et al.~\cite{Crochemore2016}.

The linear-size suffix trie of a string $T$, denoted $\LST(T) = (\VLST, \ELST)$,
is another kind of tree that represents $\Suffix(T)$,
where each edge is labeled by a single character.
The set $\VLST$ of nodes of $\LST(T)$ is a union of the two disjoint
subsets of nodes such that $\VLST = \VLSTone \cup \VLSTtwo$,
where the nodes $v_1 \in \VLSTone$ are said to be of \emph{type-1},
and the nodes $v_2 \in \VLSTtwo$ are said to be of \emph{type-2}.
We have that $\VLSTone = \VSTree$, namely,
the type-1 nodes of $\LST(T)$ represent the right-maximal substrings in $T$.
On the other hand, $\VLSTtwo$ represent the \emph{quasi} right-maximal substrings in $T$, which are defined as follows:
For a character $c \in \Sigma$ and non-empty string $u \in \Sigma^*$,
string $cu$ is called a quasi right-maximal substring in $T$
iff $u \in \RightM(T)$, $cu \in \Substr(T)$, and $cu \notin \RightM(T)$. 

Intuitively, $\LST(T)$ can be obtained from $\STrie(T)$
by keeping the three following components:
(1) All branching nodes and leaves (this gives us type-1 nodes);
(2) All non-branching nodes whose suffix links point to type-1 nodes (this gives us type-2 nodes);
(3) Only the first characters in the compressed path labels.

Formally, for each edge $(u,v)$ in $\LST(T)$,
the edge $(u,v)$ in $\LST(T)$ is labeled by the character $c = \str(v)[|\str(u)|+1]$.
If $|\str(v)|-|\str(u)| > 1$,
which means that the corresponding suffix trie path is unary and
$c$ is followed by some more character(s) in the unary path,
then a 1-bit information $\Plus$ is appended to the label $c$ of the LST edge $(u,v)$.

See the middle diagram of Figure~\ref{fig:suffix_tree_lstrie}
for an example of linear-size suffix tries.

Crochemore et al.~\cite{Crochemore2016} showed that
the number of nodes and edges in $\LST(T)$ for a string $T$ of length $n$ is $O(n)$ (Theorem 1 of~\cite{Crochemore2016}).
Below, we present tighter bounds:
\begin{lemma}\label{lem:LST_size}
  For a string $T$ of length $n \geq 3$,
  the number $|\VLST|$ of nodes in $\LST(T)$ is at most $3n-3$
  and the number $|\ELST|$ of edges in $\LST(T)$ is at most $3n-4$.
\end{lemma}
\begin{proof}
It is well known that the number of nodes of $\STree(T)$ is at most $2n-1$
for $n \geq 2$.
Thus, the number of type-1 nodes of $\LST(T)$ is at most $2n-1$ as well.
Hendrian et al.~\cite{Hendrian2019,abs-2301-04295}
pointed out that there is a one-to-one correspondence between
the type-2 nodes in $\LST(T)$ and the \emph{secondary edges}
of the directed acyclic word graph (DAWG)~\cite{Blumer1985} for the reversed string of $T$,
\sinote*{added}{%
  which are equivalent to the \emph{soft Weiner links} of $\STree(T)$.
  An explicit node $v$ of $\STree(T)$ has a soft Weiner link labeled with character $a \in \Sigma$ if $a \cdot \str(v) \in \Substr(T)$ but
  there is no explicit node representing $a \cdot \str(v)$.
  The (reversed) soft Weiner links are exactly the suffix links
  from the type-2 nodes of $\LST(T)$.
}%

Blumer et al.~\cite{Blumer1985} showed that 
there are at most $n-2$ secondary edges in the DAWG for any string $T$ of length $n \geq 3$.
This leads to a $(2n-1)+(n-2) = 3n-3$ upper bound on the number of nodes
in $\LST(T)$ for $n \geq 3$.
It is now clear that $\LST(T)$ can have at most $3n-4$ edges for $n \geq 3$.
\end{proof}

The \emph{fast links}~\cite{Crochemore2016}
are a key data structure for pattern matching and edge-label extraction on LSTs.
Fast links are basically suffix links of \emph{edges}, defined as follows:
For any non-root node $v$ of $\LST(T)$,
let $\slink^0(v) = v$ and $\slink^k(v) = \slink(\slink^{k-1}(v))$ for $1 \leq k \leq |\str(v)|$.
For a $\Plus$-edge $(u,v)$ of $\LST(T)$,
let $k > 0$ be the smallest integer such that
$\slink^k(u)$ is \emph{not} the parent of $\slink^k(v)$.
We note that there always exists such $k > 0$ for any $\Plus$-edge $(u,v)$,
since any character $c$ that occurs in $T$
is represented by a node (of type-1 or type-2) that is a child of the root.
Then, the fast link $\fastlink(u,v)$ of the edge $(u,v)$ points to the path from $\slink^k(u)$ to $\slink^k(v)$, namely, $\fastlink(u,v) = \langle \slink^k(u), \slink^k(v) \rangle$.
By the definition of type-2 nodes,
all the internal nodes in the destination path
$\langle \slink^k(u), \slink^k(v) \rangle$ are of type-2.
See also Figure~\ref{fig:fast_links} in Section~\ref{sec:sim_LST}.

\begin{example}
  See $\LST(T)$ with $T = \mathtt{abaabc\$}$ in the middle diagram of Figure~\ref{fig:suffix_tree_lstrie}.
  \begin{itemize}
  \item For $\Plus$-edge $(u, 1)$, $\fastlink(u, 1) = \langle t, 2 \rangle$ that consists of edges $(t, x)$ and $(x, 2)$.
  \item For $\Plus$-edge $(x, 2)$, $\fastlink(x, 2) = \langle s, 3 \rangle$ that consists of edges $(s, w)$, $(w, y)$, and $(y, 3)$.
  \item For $\Plus$-edges $(y, 3)$, $(u,4)$ and $(t,5)$, $\fastlink(y, 3) = \fastlink(u,4) = \fastlink(u,5) = \langle r, 6 \rangle$ that consists of edges $(r,v)$ and $(v,6)$.
  \end{itemize}
\end{example}

The following lemma shows that the entire string label of a given edge $(u,v)$ in $\LST(T)$ can be retrieved efficiently.
We present a formal proof for completeness:

\begin{lemma}[Theorem 2 of~\cite{Crochemore2016}] \label{lem:LST_edge_label_retrival}
  Given an edge $(u,v)$ of $\LST(T)$,
  the string label $x = \str(v)[|\str(u)|+1..|\str(v)|]$ of the edge $(u,v)$ can be retrieved in $O(\ell)$ time by using fast links, where $\ell = |x|$.
\end{lemma}

\begin{proof}
  We take the fast link
  of the edge $(u,v)$ and let $\fastlink(u,v) = \langle u', v' \rangle$.
  By the definition of fast links, $u'$ is not the parent of $v'$.
  Let $q_1, \ldots, q_h, q_{h+1}$ be the sequence of the nodes in the path $\langle u', v' \rangle$ in increasing order of depth,
  where $q_1$ is the child of $u'$ and $q_{h+1} = v'$.
  By the definition of fast links and type-2 nodes,
  $q_1, \ldots, q_h$ are all type-2 nodes which are all unary.
  We take the character labels of their unique out-going edges,
  which allows us to retrieve $h$ characters in the string label of the given edge $(u,v)$.
  For each $1 \leq i \leq h$,
  if the edge $(q_{i}, q_{i+1})$ is a $\Plus$-edge,
  then this means that there remain some characters to retrieve in this part of the
  string edge label for $(u,v)$,
  so we recursively apply the same procedure as above to this edge $(q_{i-1}, q_{i})$.
  Otherwise, we are done with this part of the string label of the given edge $(u,v)$.

  Each time we apply a fast link, we retrieve at least one new character
  in the string edge label of $(u,v)$.
  Thus, the number of applications to fast links is bounded by $\ell$.
  Since $\LST(T)$ is a tree,
  by traversing up from the bottom node $v'$ to the top node $u'$
  in the path $\langle u', v' \rangle$,
  one can obtain each edge in the path in $O(1)$ time independently of the alphabet size\footnote{The original algorithm \textsc{Decompact} described in~\cite{Crochemore2016} traverses the given path from top to bottom. Since the first node in the path is always of type-1, it requires $O(\log \sigma)$ time to locate the first edge in the path. This leads to an $O(\ell \log \sigma)$-time bound to retrieve the string edge label, despite of the claimed $O(\ell)$-time bound in Theorem 2 of~\cite{Crochemore2016}.}.
\end{proof}

\begin{lemma}[Theorem 2 of~\cite{Crochemore2016}] \label{lem:LST_pattern_matching}
  Given a pattern $P$ of length $m$,
  one can determine whether $P$ occurs in the text $T$ in $O(m \log \sigma)$ time
  using $\LST(T)$.
\end{lemma}
\begin{proof}
  We first traverse $P$ from the root of $\LST(T)$
  as long as we can move on with non $\Plus$-edges.
  We perform the following:
  \begin{itemize}
  \item If we find the first mismatch with a non $\Plus$-edge,
  then we are done and report that $P$ does not occur in $T$.
  \item Otherwise (if we encounter a $\Plus$-edge during the traversal), then perform the following:
    \begin{itemize}
      \item If we find a mismatch with the first character explicitly stored at one of these $\Plus$-edges,
  then we are done and report that $P$ does not occur in $T$.
      \item Otherwise, we perform the string edge label extraction algorithm
  of Lemma~\ref{lem:LST_edge_label_retrival} to the $\Plus$-edge,
  but in this case we traverse a given path in a top-down manner.
  Each time we decode a character in the edge label, compare it to the corresponding character in $P$.
  If there is a mismatch, report that $P$ does not occur in $T$.
  Otherwise, continue traversing $\LST(T)$ with $P$.
     \end{itemize}
  \end{itemize}
  We perform the above procedure until we locate the path that spells out $P$ or find the first mismatch.

  The cases without decoding $\Plus$-edges are trivial.
  Let us consider the case where we decode some $\Plus$-edges,
  and let $e_1, \ldots, e_j$ be the sequence of edges in the path that we have traversed in finding $P$.
  The total cost for retrieving the string labels for $e_1, \ldots, e_{j-1}$ (except for the last edge $e_j$) is clearly linear in the pattern length $m$ multiplied by a $\log \sigma$ factor.
  Note that the end-point of the traversal of $P$ may lie in the middle of the string label of the last edge $e_j$, and that the entire string label of $e_j$ can be much longer than $P$ (see~\cite{abs-2301-04295} for a lower bound instance).
  However, it has been shown in~\cite{Hendrian2019,abs-2301-04295} that
  the number of applications of the fast links for finding the locus of $P$
  in the last edge can be bounded by $m$.
  This is basically because the length of the suffix link chain from
  the origin node $u_j$ of the last edge $e_j = (u_j, v_j)$ is $|\str(u_j)| < m$,
  and since each character $c$ occurring in $T$ is represented by a child of the root of $\LST(T)$.
  By recursively applying the fast links in a top-down manner for a given path,
  we can locate $P$ or find the first mismatch in a total of $O(m \log \sigma)$ time.
\end{proof}

\begin{figure}[h]
  \centering
  \includegraphics[scale=0.4]{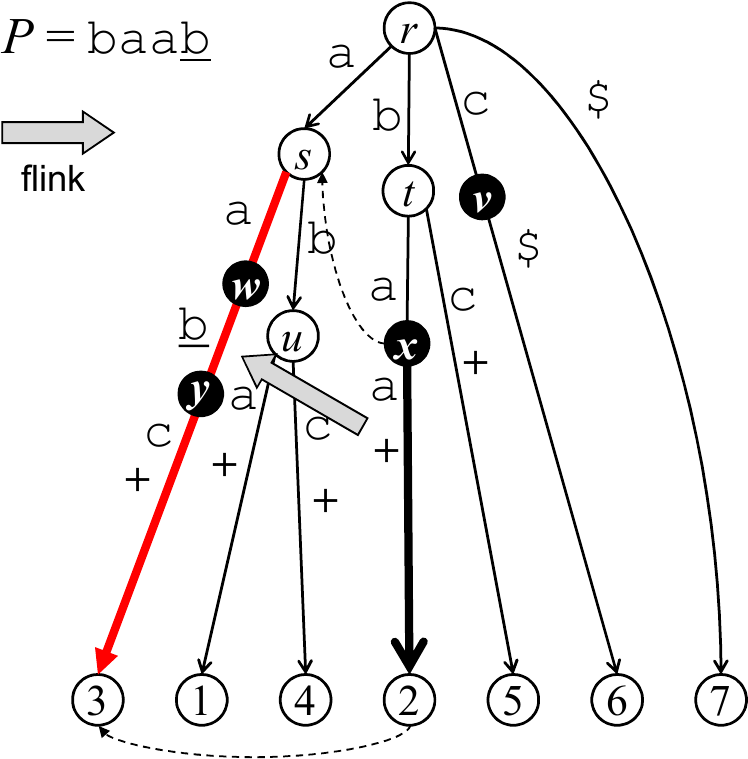}
  \caption{Matching pattern $P = \mathtt{baab}$ on $\LST(T)$ with $T = \mathtt{abaabc\texttt{\$}}$. After traversing $\mathtt{baa}$ from the root, we encounter $+$. The following character $\underline{\mathtt{b}}$ is decoded by the fast link from the bold edge $(x, 2)$ to the red bold path $\langle s, 3 \rangle$.}
  \label{fig:matching_lstrie}
\end{figure}

An example for pattern matching with $\LST(T)$ is illustrated in Figure~\ref{fig:matching_lstrie}.

\begin{remark} \label{rem:lst_occ}
Since $\LST(T)$ contains some unary paths of type-2 nodes,
a na\"ive traversal of the subtree rooted at the locus of the found pattern $P$
may take $O(\occ \cdot n)$ time, where $\occ$ is the number of pattern occurrences.
If we store the unlabeled edges of the corresponding suffix tree for all unary paths of $\LST(T)$,
then we can report all pattern occurrences in $O(\occ)$ time after locating $P$ on $\LST(T)$.
Since there are at most $n-1$ type-2 nodes,
this can require extra $n-1$ edges in the worst case.
\end{remark}

%% file: sim_lst.tex
\section{Simplified Linear-size Suffix Tries}
\label{sec:sim_LST}

In this section, we present our simplified version of
linear-size suffix tries (LSTs), called \emph{simplified linear-size suffix tries} (\emph{simplified LSTs}).

\subsection{Definition of $\simLST(T)$}

The simplified LST of a string $T$, denoted $\simLST(T) = (\VsimLST, \EsimLST)$,
is basically the LST \emph{without type-2 nodes},
except for those that are children of the root.
Namely, $\VsimLST = \VSTree \cup \Sigma_T$, where
$\Sigma_T = \{T[i] \mid 1 \leq i \leq |T|\}$ is the set of
distinct characters occurring in $T$.
For a character $c \in \Sigma_T$,
$c$ is represented by a type-1 node if $c$ is right-maximal in $T$,
and $c$ is represented by a type-2 node otherwise.
See the right diagram of Figure~\ref{fig:suffix_tree_lstrie}
for an example of simplified LSTs.

\begin{figure}[tb]
	\centering
        \includegraphics[scale=0.5]{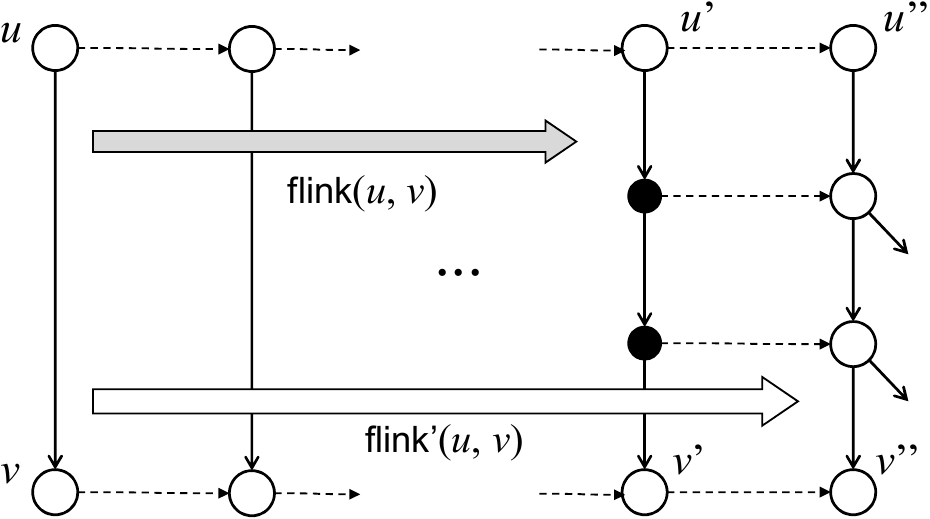}
	\caption{Illustration for the fast link $\fastlink(u, v) = \langle u', v' \rangle$ on $\LST(T)$ and the modified fast link $\fastlink'(u, v)$ on $\simLST(T) = \langle u'', v'' \rangle$ for a $\Plus$-edge $(u, v)$, where $u'' = \slink(u')$ and $v'' = \slink(v')$.}
	\label{fig:fast_links}
\end{figure}

Now we define the modified fast links 
in a similar way to Section~\ref{sec:lst}, but using the paths on $\simLST(T)$:
Let $(u, v)$ be any $\Plus$-edge of $\simLST(T)$.
Let $k > 0$ be the smallest integer such that
$\slink^k(u)$ is \emph{not} the parent of $\slink^k(v)$.
Then, the modified fast link $\fastlink'(u,v)$ of the $\Plus$-edge $(u,v)$
is defined by $\fastlink'(u,v) = \langle \slink^k(u), \slink^k(v) \rangle$.
See also Figure~\ref{fig:fast_links} for illustration.

Intuitively, our modified fast link $\fastlink'(u, v)$ on $\simLST(T)$ moves up
one more step further in the suffix link chain
than the original fast link $\fastlink(u, v)$ on $\LST(T)$.

\begin{example}
  See $\simLST(T)$ with $T = \mathtt{abaabc\$}$ in the right diagram of Figure~\ref{fig:suffix_tree_lstrie}.
  \begin{itemize}
  \item For $\Plus$-edges $(u,1)$ and $(t, 2)$, $\fastlink'(u, 1) = \fastlink'(t, 2) = \langle r, 3 \rangle$ that consists of edges $(r, s)$ and $(s, 3)$.
  \item For $\Plus$-edge $(s, 3)$, $\fastlink'(s, 3) = \langle r, 4 \rangle$ that consists of edges $(r, s)$, $(s, u)$, and $(u, 4)$.
  \item For $\Plus$-edges $(u,4)$ and $(t,5)$, $\fastlink'(u,4) = \fastlink'(u,5) = \langle r, 6 \rangle$ that consists of edges $(r, v)$ and $(v, 6)$.
  \end{itemize}
\end{example}

\subsection{Size Bounds for $\simLST(T)$}

In this subsection, we present a tight bound for the size of $\simLST(T)$.

\sinote*{added}{%
  When we do \emph{not} have an end-marker $\$$ at the right end of $T$,
  then some suffixes of $T$ may not be represented by leaves
  in the suffix tree (and thus in the linear-size suffix tries as well).
  Let $s_1, \ldots, s_k$ be the suffixes of $T$
  each occurring at least twice in $T$, arranged in decreasing order of lengths.
  Let $s_1, \ldots, s_j$ be the repeating suffixes of $T$
  that have a unique character $a \in \Sigma$ such that $s_ia \in \Substr(T)$
  for $1 \leq i \leq j \leq k$.
  The original suffix tree proposed by Wiener~\cite{Weiner1973}
  has non-branching internal explicit nodes for all such
  $j$ suffixes $s_1, \ldots, s_j$ of $T$,
  and this version of the suffix tree is sometimes called the \emph{Weiner tree}.
  On the other hand, in the version of the suffix tree proposed by Ukkonen~\cite{Ukkonen1995}, sometimes referred to as the \emph{Ukkonen} tree,
  the suffixes $s_1, \ldots, s_j$ are represented by implicit nodes.
  Thus, the Weiner tree has $j$ more nodes than the Ukkonen tree.
  It is clear that Weiner trees and Ukkonen trees coincide
  when the input strings terminate with $\$$.

The next lemma deals with the case where $\simLST(T)$ is defined with the Weiner tree without an assumption that $T$ terminates with $\$$, where there are $j$~($j \leq k$) more nodes than the Ukkonen tree.
The case where $T$ terminates with $\$$ is obtained by simply setting $k = j = 0$.
}%

\begin{lemma} \label{lem:simLST_size}
  For any string $T$ of length $n$,
  $|\VsimLST| \leq 2n$ and $|\EsimLST| \leq 2n-1$,
  and theses bounds are tight.
\end{lemma}

\begin{proof}
  We first revisit the upper bound for the number of nodes in $\STree(T)$.
  Let $k$ be the number of suffixes of $T$ which are not represented by leaves of $\STree(T)$.
  Then there are at most $k$ non-branching type-1 nodes,
  and there are exactly $n-k$ leaves in $\STree(T)$.
  Then the maximum number of branching nodes in $\STree(T)$ is $n-k-1$,
  which can be achieved only if each branching node has exactly two children.
  However, the root must have exactly $\sigma_T$ children.
  This implies that the number of nodes in $\STree(T)$,
  which is the same as the number of type-1 nodes in $\simLST(T)$,
  is at most $k + (n-k) + (n-k-1) - (\sigma_T - 2) = 2n - k - \sigma_T + 1$ for $\sigma_T \geq 2$.
  The last character $T[n]$ is always right-maximal, so it is represented by a type-1 node,
  and hence the number of type-2 nodes is at most $\sigma_T-1$.
  In total, there are at most $(2n -k - \sigma_T + 1) + (\sigma_T-1) = 2n-k \leq 2n$ nodes, and at most $2n-1$ edges in $\simLST(T)$ for $\sigma_T \geq 2$.
  For $\sigma_T = 1$, there are only $n+1$ nodes and $n$ edges.
  
  A matching lower bound can be obtained e.g. when all the characters in $T$ are distinct, i.e., $T = c_1 \cdots c_n$ with $\Sigma_T = \{c_1, \ldots, c_n\}$.
  The only branching node in $\simLST(T)$ is the root,
  and the root has exactly $\sigma_T-1 = n - 1$ type-2 children
  since $c_n$ is right-maximal (represented by a leaf)
  and any other $c_i$~($1 \leq i < n$) are not right-maximal.
  Since there are exactly $n$ leaves,
  we have $1 + n-1 + n = 2n$ nodes and $2n-1$ edges.
\end{proof}

\subsection{Label Extraction and Pattern Matching with $\simLST(T)$}

The following lemma states that we can basically use the same
edge label extraction algorithm in $\simLST(T)$
as the case of $\LST(T)$.

\begin{lemma} \label{lem:sim_LST_edge_label_retrival}
  Given an edge $(u,v)$ of $\simLST(T)$,
  its label $x = \str(v)[|\str(u)|+1..|\str(v)|]$ can be retrieved in $O(\ell)$ time by using modified fast links, where $\ell = |x|$.
\end{lemma}

\begin{proof}
  Recall that the edge label extraction algorithm of Lemma~\ref{lem:LST_edge_label_retrival} for $\LST(T)$ traverses the path $\langle u', v' \rangle = \fastlink(u, v)$ in a bottom-up manner.
  Although our modified fast link for a given edge arrives
  at a non-branching path $\mathcal{P}$,
  since $\LST(T)$ is a rooted tree,
  we can obviously obtain each edge in the path $\mathcal{P}$
  in $O(1)$ time in a bottom-up manner.
  This shows the lemma.
\end{proof}

Next, we show how to perform pattern matching with $\simLST(T)$.
For this sake, we perform a standard $O(n)$-time DFS on $\simLST(T)$
so that, given a pair of nodes $u, v$ in $\simLST(T)$,
one can determine whether $u$ is an ancestor of $v$ or not in $O(1)$ time.

\begin{lemma} \label{lem:sim_LST_pattern_matching}
  Given a pattern $P$ of length $m$,
  one can find all $\occ$ occurrences of $P$ in the text $T$ in $O(m \log \sigma + \occ)$ time
  using $\simLST(T)$.
\end{lemma}

\begin{proof}
  Recall that the tree topology of $\simLST(T)$ is almost identical to $\STree(T)$
  except for the non-branching children of the root.
  Thus, once locating $P$ on $\simLST(T)$,
  one can report all $\occ$ occurrences in $O(\occ)$ time by a standard traversal
  of the subtree as in the case of suffix trees.
  Below, we show how to determine whether $P$ occurs in $T$,
  and if so, find the locus for $P$ on $\simLST(T)$.
  
  Recall that the edge-label extraction algorithm
  on $\LST(T)$ described in Lemma~\ref{lem:LST_pattern_matching}
  traverses the paths obtained by following fast links
  in a \emph{top down} manner.
  We have the two following cases to consider:

  When a given pattern $P$ is a substring of the text $T$,
  then a direct application of Lemma~\ref{lem:LST_pattern_matching} for
  $\LST(T)$ to our $\simLST(T)$ works correctly
  within the $O(m \log \sigma)$-time bound.

  When $P$ is not a substring of $T$,
  then a direct application of Lemma~\ref{lem:LST_pattern_matching} for
  $\LST(T)$ to our $\simLST(T)$ may report a false positive.
  This is because of our modified fast links that arrive at
  non-branching paths, and because of the following suffix-property:
  Suppose we are to extract the label of edge $(u, v)$,
  and let $\fastlink'(u, v) = \langle u', v' \rangle$.
  Assume that $P$ has the first mismatch on the edge $(u, v)$ (see also Figure~\ref{fig:false_positive} for illustration).
  Recall that $S = \str(u')$ is a proper suffix of
  $Q = \str(u) = P[1..|\str(u)|]$,
  since $u'$ is on the suffix link chain from $u$ towards the root.
  Let $U$ be the string such that $QU$ is the shortest
  prefix of $P$ that is not a substring of $T$.
  Since $S$ is a proper suffix of $Q$,
  it is possible that $SU$ is a substring of $T$.
  In such cases, the direct application of Lemma~\ref{lem:LST_pattern_matching}
  can report a false positive that non-substring pattern $P$ is a substring of $T$ on $\simLST(T)$.

\begin{figure}[h]
	\centering
        \includegraphics[scale=0.48]{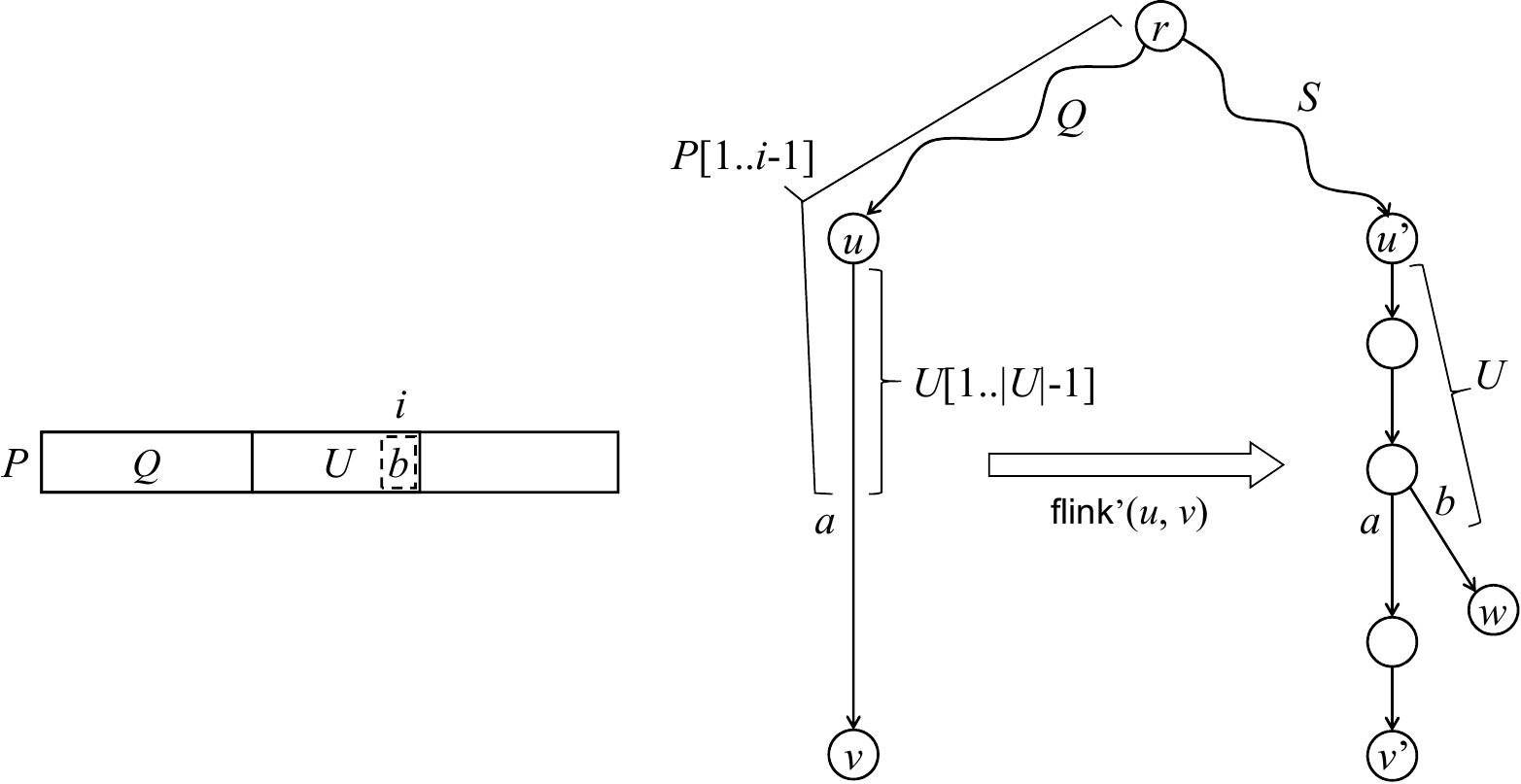}
	\caption{Illustration showing false positives that may occur in a direct application of the top-down pattern matching algorithm with $\LST(T)$ to $\simLST(T)$. $P[i] = b \neq a$ is the first mismatch found on edge $(u, v)$ in the search for $P$ from the root $r$. The modified fast link $\fastlink'(u,v)$ points to the path $\langle u', v' \rangle$ and there can be a path spelling out $U$ from $u'$. However, such a false positive occurs iff the destination node $w$ of $SU$ is \emph{not} an ancestor of $v'$.}
	\label{fig:false_positive}
\end{figure}
  
  The above false positive can be avoided as follows:
  Let $(x, y)$ be any edge of which label we are to extract for pattern matching with $P$
  on $\simLST(T)$.
  Let $\langle x', y' \rangle = \fastlink'(x, y)$.
  We continue extracting the edge label for $(x', y')$
  if $x'$ is an ancestor of $y'$,
  and we immediately stop extracting edge labels
  if $x'$ is not an ancestor of $y'$.
  This allows for performing pattern matching correctly without reporting false positives.

  The time complexity for locating $P$ remains $O(m \log \sigma)$ since
  we can determine whether a given node is an ancestor of another given node
  in constant time, after a standard linear-time preprocessing on the tree.
  We can then report all $\occ$ occurrences in $O(\occ)$ time by simply navigating the subtree under the locus of $P$, since all the type-2 nodes are of string depth 1.  
\end{proof}

\begin{figure}[h]
  \centering
  \includegraphics[scale=0.4]{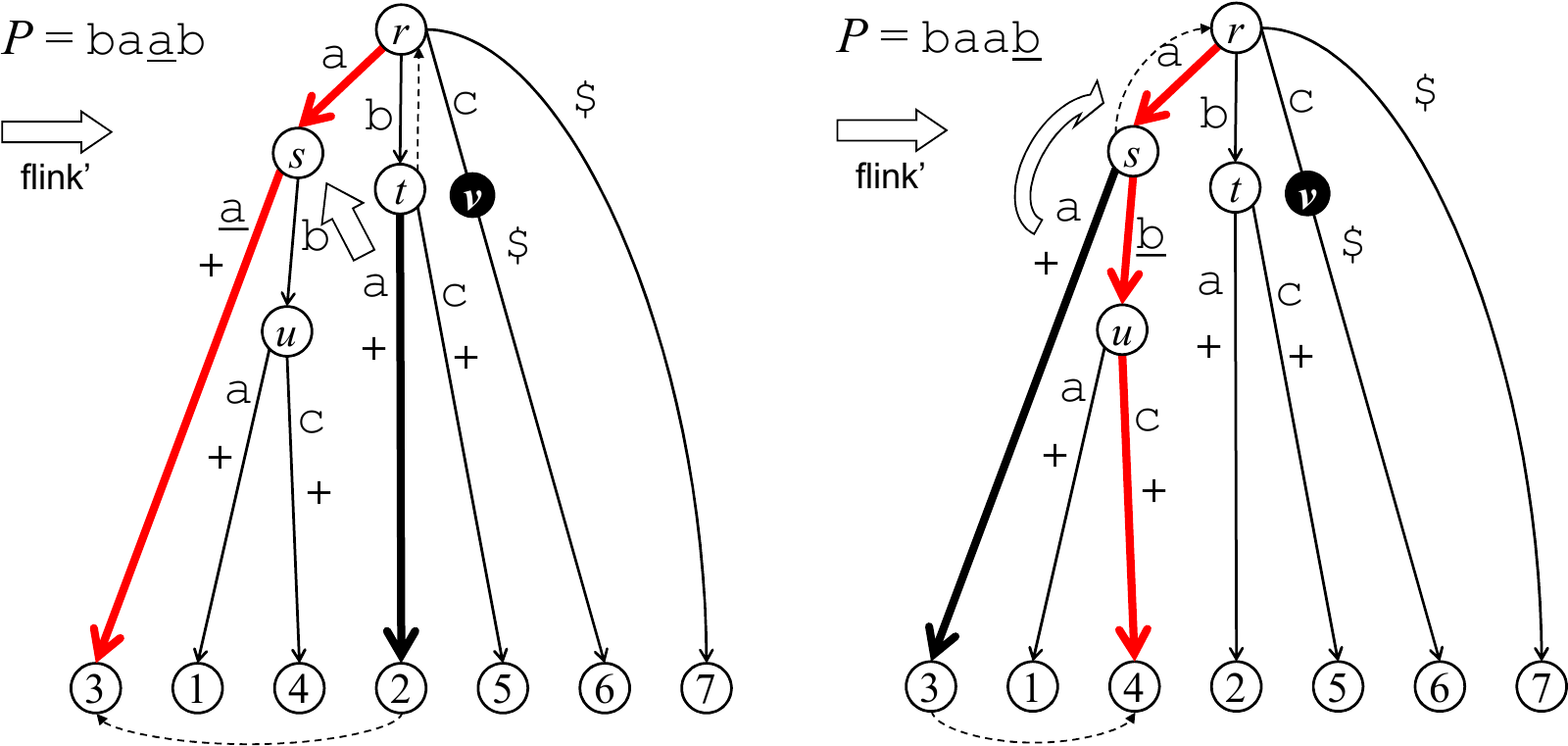}
  \caption{Matching pattern $P = \mathtt{baab}$ on $\simLST(T)$ with $T = \mathtt{abaabc\texttt{\$}}$. After traversing $\mathtt{ba}$ from the root, we encounter $+$ (the left diagram). The following character $\underline{\mathtt{a}}$ is decoded by the modified fast link from the bold edge $(t, 2)$ to the red bold path $\langle s, 3 \rangle$. We again encounter $+$ on edge $(s, 3)$ (the right diagram). The following character $\underline{\mathtt{b}}$ is decoded by the modified fast link from the bold edge $(s, 3)$ to the red bold path $\langle r, 4 \rangle$.}
  \label{fig:matching_sim_lstrie}
\end{figure}

Figure~\ref{fig:matching_sim_lstrie} illustrates an example of pattern matching with $\simLST(T)$, where the pattern $P$ occurs in $T$.

\begin{figure}[h]
  \centering
  \includegraphics[scale=0.4]{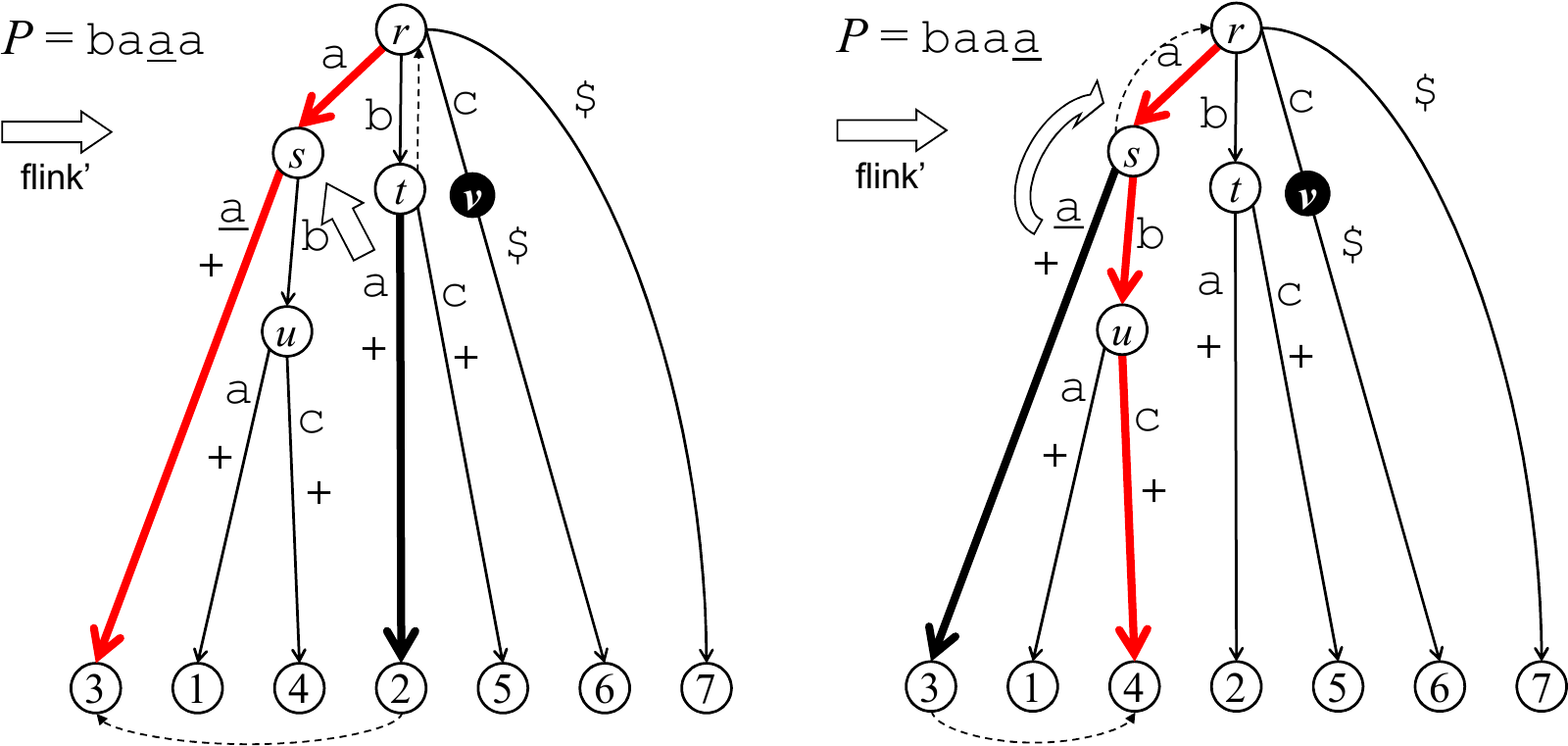}
  \caption{Matching pattern $P = \mathtt{baaa}$ on $\simLST(T)$ with $T = \mathtt{abaabc\texttt{\$}}$. It is the same as in Figure~\ref{fig:matching_sim_lstrie} up to $P[1..3] = \mathtt{baa}$, and we encounter $+$ on edge $(s, 3)$ (the right diagram). The following character $\underline{\mathtt{a}}$ is detected from node $s$. However, since it is not on the red path, $P = \mathtt{baaa}$ does not occur in $T$.}
  \label{fig:failed_matching_sim_lstrie}
\end{figure}

Figure~\ref{fig:failed_matching_sim_lstrie} illustrates an example of pattern matching with $\simLST(T)$, where the pattern $P$ does not occur in $T$.

\subsection{Construction of $\simLST(T)$}

\begin{lemma}
  Given a string $T$ of length $n$ over an integer alphabet of polynomial size in $n$,
  one can construct $\simLST(T)$ in $O(n)$ time and space.
\end{lemma}

\begin{proof}
  Fujishige et al.~\cite{FujishigeTIBT23} showed that
  $\LST(T)$ augmented with fast links can be built
  in $O(n)$ time and space in the case of integer alphabets.

  For each edge $(u, v)$ of $\LST(T)$ such that both $u$ and $v$ are type-1 nodes,
  let $\langle u', v' \rangle = \fastlink(u, v)$.
  Then, we create our modified fast link for edge $(u, v)$ as
  $\fastlink'(u, v) = \langle \slink(u'), \slink(v') \rangle$.
  \begin{itemize}
    \item[(1)] If $u'$ is not the root, then we create a $\Plus$-edge $(u', v')$ by discarding all type-2 nodes
  in the original path $\langle u', v' \rangle$.
  We then set $\fastlink'(u', v') = \langle \slink(u'), \slink(v') \rangle$.


    \item[(2)] If $u'$ is the root, then its child $w$ in the path is a type-2 (non-branching) node. We perform the same procedures as Case (1) to the path $\langle w, v \rangle$. We then keep the non-branching type-2 node $w$ in our $\simLST(T)$.
  \end{itemize}
  All of the above operations can performed in $O(n)$ total time,
  since there are $O(n)$ total nodes and edges in $\LST(T)$ and $\simLST(T)$.  
\end{proof}

%% file: lcdawg.tex
\section{Linear-size CDAWGs} \label{sec:lcdawg}

This section reviews 
the \emph{linear-size CDAWG} (\emph{LCDAWG}) data structure,
originally proposed by Takagi et al.~\cite{Takagi2017}.

The LCDAWG of a string $T$, denoted $\LCDAWG(T) = (\VLCDAWG, \ELCDAWG)$,
is another kind of DAG that represents $\Suffix(T)$,
where each edge is labeled by a single character.
The set $\VCDAWG$ of nodes of $\LCDAWG(T)$ is a union of the two disjoint
subsets of nodes such that $\VLCDAWG = \VLCDAWGone \cup \VLCDAWGtwo$,
where the nodes $v_1 \in \VLCDAWGone$ are said to be of \emph{type-1},
and the nodes $v_2 \in \VLCDAWGtwo$ are said to be of \emph{type-2}.
We have that $\VLCDAWGone = \VCDAWG$, namely,
the type-1 nodes of $\LCDAWG(T)$ represent the maximal substrings in $T$.
On the other hand, $\VLCDAWGtwo$ represent the \emph{quasi} maximal substrings in $T$, which are defined as follows:
For a character $c \in \Sigma$ and non-empty string $u \in \Sigma^*$,
string $cu$ is called a quasi maximal substring in $T$
iff $u \in \M(T)$, $cu \in \Substr(T)$, and $cu \notin \M(T)$.
Each edge of $\LCDAWG(T)$ that represents a substring of length $\geq 2$
stores the first character of its string label
and the one-bit information $\Plus$.
Each edge of length $1$ stores only the first character of its string label.

See the middle diagram of Figure~\ref{fig:cdawg_lcdawg}
for an example of LCDAWGs.

The following lemma is immediate from the definition of $\LCDAWG(T)$:
\begin{lemma}[Adapted from Lemma 1 of~\cite{Takagi2017}] \label{lem:LCDAWG_size}
For any string $T$,
\begin{itemize}
  \item The number of type-1 nodes in $\LCDAWG(T)$ is $|\VCDAWG|$.
  \item The number of type-2 nodes in $\LCDAWG(T)$ is at most $\lefte(T)$.
  \item The number of all edges in $\LCDAWG(T)$ is at most $\righte(T)+\lefte(T)$.
\end{itemize}
\end{lemma}

The fast link of each $\Plus$-edge $(u, a+, v)$ of $\LCDAWG(T)$
is defined as follows:
Let $x$ be the string label for this edge $(u, a+, v)$ in the original $\CDAWG(T)$.
Then, $\fastlink(u, a+, v) = \langle u'_1, \ldots, u'_k, v' \rangle$ where
\begin{itemize}
  \item the path $\langle u'_1, \ldots, u'_k, v' \rangle$ spells out $x$, and
  \item $\str(u'_1)$ is the longest proper suffix of $\str(u)$
such that the path $\langle u'_1, \ldots, u'_k, v' \rangle$
contains at least three nodes including $u'_1$ and $v'$ (namely $k \geq 2$).
\end{itemize}
Note that all the inner nodes $u'_2, \ldots, u'_k$ in the path $\langle u'_1, \ldots, u'_k, v' \rangle$ are type-2 nodes.
Also, by the suffix property, $\str(v')$ is a suffix of $\str(v)$.

\begin{example}
  See $\LCDAWG(T)$ with $T = \mathtt{abaabc\$}$ in the middle diagram of Figure~\ref{fig:cdawg_lcdawg}.
  \begin{itemize}
  \item For $\Plus$-edge $(x, \mathtt{a}+, z)$, $\fastlink(x, \mathtt{a}+, z)$ points to the path that consists of edges $(s, \mathtt{a}, w)$, $(w, \mathtt{b}, y)$, and $(y, \mathtt{c}+, z)$.
  \item For $\Plus$-edges $(y, \mathtt{c}+, z)$ and $(u, \mathtt{c}+, z)$, $\fastlink(y, \mathtt{c}+, z) = \fastlink'(u, \mathtt{c}+, z)$ both point to the path that consists of edges $(r, \mathtt{c}, v)$ and $(v, \mathtt{\$}, z)$.
  \end{itemize}
\end{example}

Since $\LCDAWG(T)$ is a DAG,
a na\"ive extension of the bottom-up approach for edge label extraction requires super-linear space.
This is because
explicitly storing the destination path $\langle u'_1, \ldots, u'_k, v' \rangle$
for each fast link $\fastlink(u, v)$ requires $O(k)$ space and $k$ cannot be bounded by a constant.
Takagi et al.~\cite{Takagi2017} employed a top-down approach for edge label extraction
on $\LCDAWG(T)$, following the original top-down approach on $\LST(T)$ by Crochemore et al.~\cite{Crochemore2016}.
In order to extract the edge label in an online manner (from the first character to the last one),
Takagi et al.~\cite{Takagi2017} represent the fast links
by a \emph{context free grammar}
\sinote*{added}{of size $\righte(T)+\lefte(T)$},
where each $\Plus$-edge corresponds to a non-terminal,
and each non $\Plus$-edge corresponds to a terminal.
Each fast link $\fastlink(u, v) = \langle u'_1, \ldots, u'_k, v' \rangle$
induces its corresponding production of which
the left-hand side is the origin edge $(u, v)$
and the right-hand is the concatenation of the non-terminals and terminals
for the edges in the path $\langle u'_1, \ldots, u'_k, v' \rangle$.
Then, the decompressed string that is derived by the production is the string label of the edge $(u, v)$.

For efficient pattern matching, it is important to be able to access
the characters of the decompressed string in a sequential manner (from the first to the last).
For this sake, Takagi et al.~\cite{Takagi2017} enhanced their context-free grammar
with the algorithm of Gasieniec et al.~\cite{GasieniecKPS05}
that allows for $O(\ell)$-time sequential access
for the label of a given edge of length $\ell$,
after linear-time preprocessing \sinote*{modified}{in the size of the grammar}.

Overall, Takagi et al.~\cite{Takagi2017} showed the following:
\begin{theorem}[Adapted from Theorem 9 of~\cite{Takagi2017}]
Let $T$ be an input string of length $n$ over an alphabet of size $\sigma$.
$\LCDAWG(T)$ occupies $O(\righte(T)+\lefte(T))$ space,
supports pattern matching queries in $O(m \log \sigma + \occ)$ time
for a given pattern of length $m$,
and can be built in $O((\righte(T)+\lefte(T)) \log \sigma)$ time
with $O(\righte(T)+\lefte(T))$ working space if the edge sorted $\CDAWG(T)$ is given.
\end{theorem}

\begin{figure}[h]
  \centering
  \includegraphics[scale=0.4]{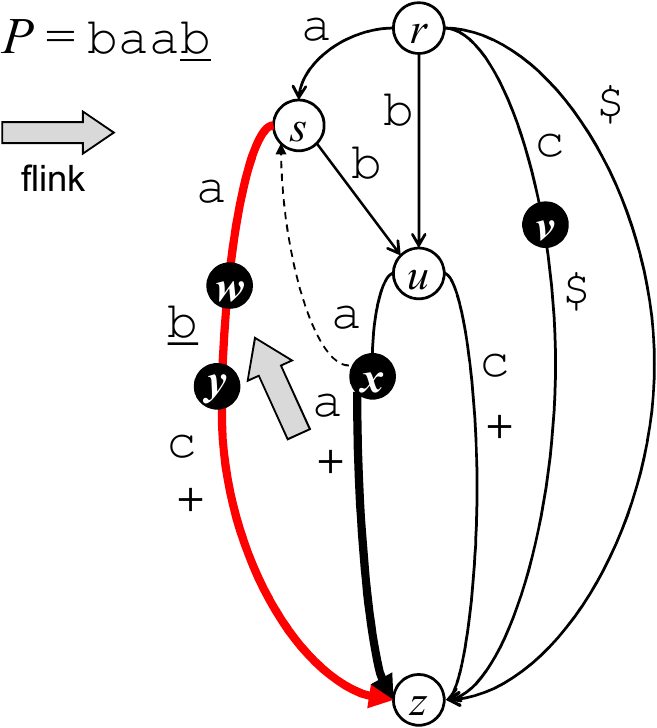}
  \caption{Matching pattern $P = \mathtt{baab}$ on $\LCDAWG(T)$ with $T = \mathtt{abaabc\texttt{\$}}$. After traversing $\mathtt{baa}$ from the source, we encounter $+$. The following character $\underline{\mathtt{b}}$ is decoded by the fast link from the bold edge $(x, z)$ to the red bold path $\langle s, w, y, z \rangle$.}
  \label{fig:matching_lcdawg}
\end{figure}

An example for pattern matching with $\LCDAWG(T)$ is illustrated in Figure~\ref{fig:matching_lcdawg}.

%% file: sim_lcdawg.tex
\section{Simplified Linear-size CDAWG}
\label{sec:sim_LCDAWG}

In this section, we present our simplified version of
linear-size compact directed acyclic word graphs, called \emph{simplified LCDAWGs}.
Our basic idea is to remove all type-2 nodes from LCDAWGs,
except for those which are the children of the source.
However, since (L)CDAWGs are DAGs, it is non-trivial to apply
our method for the simplified LSTries to the case of (L)CDAWGs.
To overcome this difficulty, we work on some specialized tree over the CDAWG.
The details will be discussed in the following subsections.

This section will show the following theorem:
\begin{theorem} \label{theo:sim_LCDAWG}
For an input string $T$ over an alphabet of size $\sigma$,
$\simLCDAWG(T)$ occupies $O(\righte(T))$ space,
supports pattern matching queries in $O(m \log \sigma + \occ)$ time
for a given pattern of length $m$,
and can be built in $O(\righte(T) \log \sigma)$ time
with $O(\heightSLT(T))$ working space if the edge sorted $\CDAWG(T) = (\VCDAWG, \ECDAWG)$ and the string $T$ are given, where $\heightSLT(T) \leq |\VCDAWG|$ is the height of the suffix link tree of $\CDAWG(T)$.
\end{theorem}

\subsection{Definition and Space Efficiency of $\simLCDAWG(T)$}
The simplified linear-size CDAWG of a string $T$, denoted 
$\simLCDAWG(T) = (\VsimLCDAWG, \EsimLCDAWG)$,
is basically the LCDAWG \emph{without type-2 nodes},
except for those that are children of the source.
Due to this simplification, the number of nodes in $\simLCDAWG(T)$
is at most $|\VCDAWG|+\sigma_T$ and the number of edges in $\simLCDAWG(T)$
is at most $\righte(T) + \sigma_T$.
We will provide with a full description of this later in this section.

Recall that the number of type-2 nodes in $\LCDAWG(T)$ is $O(\lefte(T))$ (Lemma~\ref{lem:LCDAWG_size}) and the number of all edges in $\LCDAWG(T)$ is $O(\lefte(T)+\righte(T))$.
The following lemma shows that removing the $\lefte(T)$ factor
from the space complexity can achieve $\Omega(\sqrt{n})$ improvements in size for some strings:
\begin{lemma}[\cite{Karkkainen17}] \label{lem:lowerbound_el_er}
  There exists a family of strings $S$ of length $n$ such that
  $\lefte(S) = \Theta(n)$ and $\righte(S) = \Theta(\sqrt{n})$.
\end{lemma}

\begin{proof}
  Consider the alphabet $\Sigma = \{\mathtt{1}, \mathtt{2}, \mathtt{3}, \mathtt{4}, \ldots, \mathtt{k}\} \cup \{\overline{\mathtt{1}}, \overline{2}, \overline{3}, \overline{4}, \ldots, \overline{\mathtt{k}}, \overline{\mathtt{k+1}}\}$ of size $2k+1$ and the following string $S$ over $\Sigma$:
\[ 
S =
\overline{\mathtt{1}} 
\ \mathtt{1} \ \overline{\mathtt{2}} 
\ \mathtt{12} \ \overline{\mathtt{3}}
\ \mathtt{123} \ \overline{\mathtt{4}}
\ \mathtt{1234} \ \overline{\mathtt{5}}
\cdots \overline{\mathtt{k-1}}
\ \mathtt{1234 \cdots k-1} \ \overline{\mathtt{k}}
\ \mathtt{1234 \cdots (k-1) k} \ \overline{\mathtt{k+1}}.
\]
Note that $\M(S) = \{\varepsilon, \mathtt{1}, \mathtt{12}, \mathtt{123}, \mathtt{1234}, \ldots, \mathtt{1234 \cdots k-1}, S\}$.
For $\mathtt{1} \cdots \mathtt{i} \in \M(S) \setminus \{\varepsilon, S\}$
that corresponds to an internal node of $\CDAWG(T)$,
it has exactly two right extensions $\mathtt{i+1}$ and $\overline{\mathtt{i+1}}$,
while it has exactly $k-i$ left extensions $\overline{\mathtt{i}}, \ldots, \overline{\mathtt{k-1}}$.
Since $|\M(S) \setminus \{\varepsilon, S\}| = k-1$,
we have that
\begin{eqnarray*}
\righte(S) & = & 2k+1 + 2(k-1) = 4k-1 = \Theta(k), \\
\lefte(S) & = & 2k+1 + \sum_{1 \leq i \leq k-1}(k-i) = \Theta(k^2),
\end{eqnarray*}
where the $2k+1$ term is from all the characters in $\Sigma$
which are the left/right extensions of $\varepsilon$ (the source of $\CDAWG(S)$).
Since $|S| = \Theta(k^2)$,
we have that $\lefte(S) = \Theta(n)$ and $\righte(S) = \Theta(\sqrt{n})$.
\end{proof}

Our idea for an $O(\righte(T))$-size representation
of the CDAWG is to extend the spanning tree
$\LPT(T)$ of $\CDAWG(T)$ which consists only of the primary edges of $\CDAWG(T)$.
Consider the tree that is obtained by adding
to each node of $\LPT(T)$ all its secondary out-going edges.
In other words, this tree is obtained from $\CDAWG(T)$ by separating each
secondary edge of a node from its destination node.
We then add the type-2 children of the source to the tree, if such nodes exist.
The resulting tree is denoted by $\LPTplus(T)$.
Note that there is a trivial one-to-one correspondence between
the edges of $\LPTplus(T)$ and the edges of $\simLCDAWG(T)$.
It is thus clear that the number of edges in $\LPTplus(T)$ is
$|\EsimLCDAWG| \leq |\ECDAWG| + \sigma_T = O(\righte(T))$.
We sometimes identify the edges of $\simLCDAWG(T)$
with the edges of $\LPTplus(T)$ when there occur no confusions.

\begin{figure}[t]
	\centering
	\begin{minipage}[t]{0.49\hsize}
		\centering
		\includegraphics[scale=0.4]{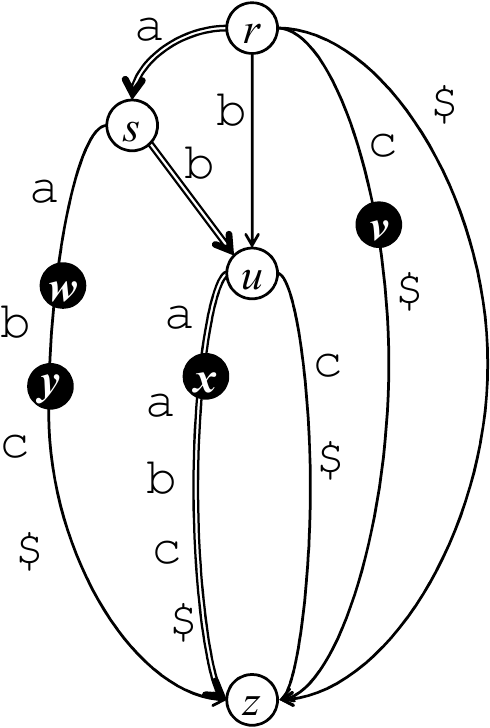}\\
		\ \small{$\LCDAWG(T)$ with string labels}
	\end{minipage} 
	\begin{minipage}[t]{0.49\hsize}
		\centering
		\includegraphics[scale=0.4]{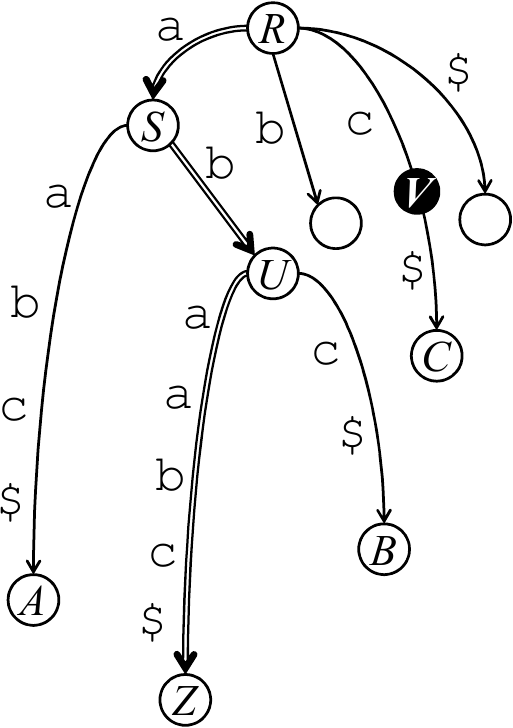}\\
		\ \small{$\LPTplus(T)$ with string labels}
	\end{minipage} \vspace*{5mm} \\ 
        	\begin{minipage}[t]{0.49\hsize}
		\centering
		\includegraphics[scale=0.4]{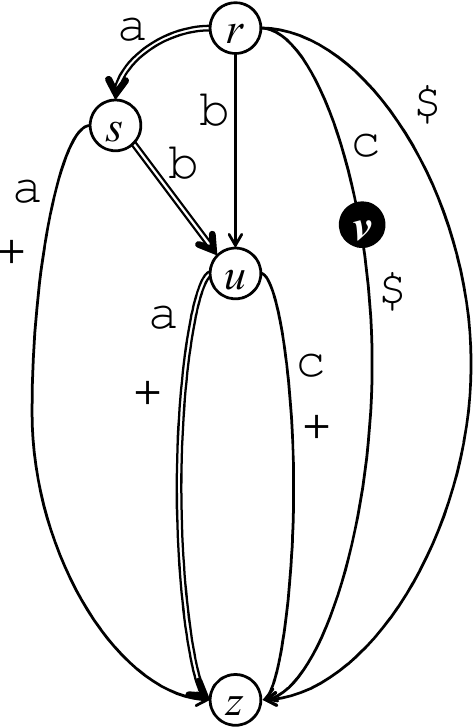}\\
		\ \small{$\simLCDAWG(T)$}
	\end{minipage}
	\begin{minipage}[t]{0.49\hsize}
		\centering
		\includegraphics[scale=0.4]{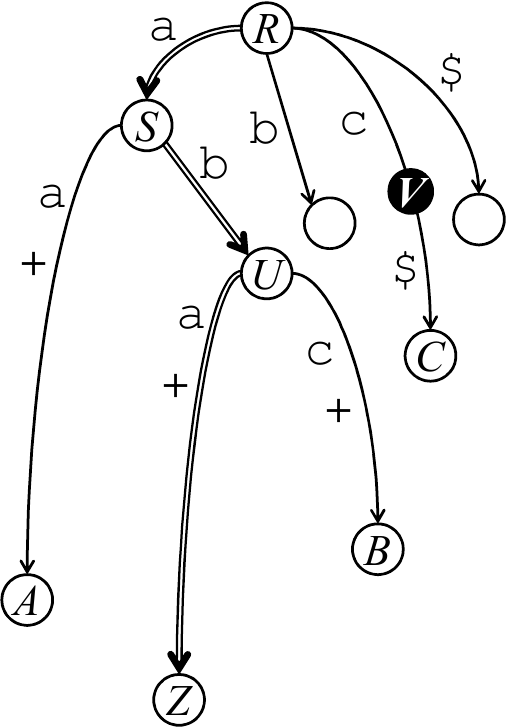}\\
		\ \small{$\LPTplus(T)$}
	\end{minipage}
	\caption{
	  Data structures for string $T = \mathtt{abaabc\texttt{\$}}$. Upper: $\LCDAWG(T)$ and $\LPTplus(T)$ with string labels. Lower: $\simLCDAWG(T)$ and $\LPTplus(T)$. The doubly-lined arcs represent the primary edges, and the singly-lined arcs represent the secondary edges.
	}
	\label{fig:lcdawg_lpt_plus}
\end{figure}

Our idea for removing the $\lefte(T)$ term from the space complexity
in the LCDAWG data structure is to apply our algorithm for $\simLST(T)$ to $\LPTplus(T)$.
Now, we obtain the following lemma:

\begin{lemma} \label{lem:LPT_path}
  Let $u_0, u_1, \ldots, u_{k-1}, u_k$ be any unary path of $\LCDAWG(T)$ such that
  both of the terminal nodes $u_0$ and $u_k$ are type-1 nodes
  (and hence all the other inner nodes $u_1, \ldots, u_{k-1}$ are type-2 nodes).
  If $u_0$ is not the source of $\LCDAWG(T)$,
  then all the edges in the path $\langle \slink(u_{0}), \slink(u_1), \ldots, \slink(u_{k-1}) \rangle$ are primary edges.
  If $u_0$ is the source of $\LCDAWG(T)$,
  then all the edges in $\langle \slink(u_1), \ldots, \slink(u_{k-1}) \rangle$ are primary edges.
\end{lemma}

\begin{figure}[t]
	\centering
        \includegraphics[scale=0.5]{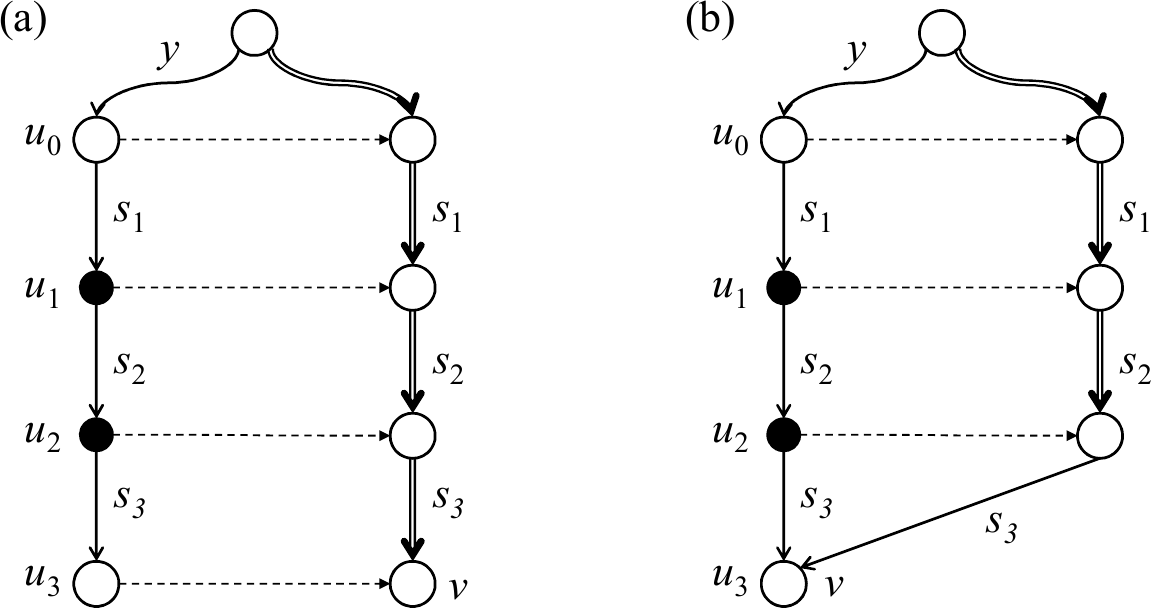}
	\caption{Illustration for the proof of Lemma~\ref{lem:LPT_path}. The edges in the path from $\slink(u_0)$ to $\slink(u_{k-1})$ spelling out $s_1 \cdots s_{k-1}$ are all primary. The last edge from $\slink(u_{k-1})$ to $v$ spelling out $s_k$ is primary in Case (a), and is secondary in Case (b).}
	\label{fig:modified_fastlinks}
\end{figure}

\begin{proof}
  We first consider the case where $u_0$ is not the source.
  See also Figure~\ref{fig:modified_fastlinks} for illustration.
  Let $y$ be the \emph{shortest} string that is represented by the node $u_0$.
  Let $x_i = \str(\slink(u_i))$ and $s_i = \str(u_{i-1}, u_i) = \str(\slink(u_{i-1}), \slink(u_i))$ for each $1 \leq i \leq k-1$.
  We prove our claim by induction on $i$:
  \begin{itemize}
  \item For $i = 1$: By the definition of suffix links, $\str(\slink(u_0))$ is the longest proper suffix of $y$, and hence $\str(\slink(u_1)) = \str(\slink(u_0)) \cdot s_1$ is the longest proper suffix of $ys_1$. Thus, there cannot be a longer path from the source that ends at $\slink(u_1)$ than the one that terminates with edge $(\slink(u_{0}), \slink(u_1))$. Therefore, $(\slink(u_{0}), \slink(u_1))$ is a primary edge.
  \item For $i \geq 2$: Suppose that $(\slink(u_{0}), \slink(u_1)), \ldots, \slink(u_{i-2}, u_{i-1})$ are all primary edges. This means that $\str(\slink(u_i)) = \str(\slink(u_0)) \cdot s_1 \cdots s_i$ is the longest proper suffix of $ys_1 \cdots s_i$. Thus, there cannot be a longer path from the source that ends at $\slink(u_i)$ than the one that terminates with edge $(\slink(u_{i-1}), \slink(u_i))$. Therefore, $(\slink(u_{i-1}), \slink(u_i))$ is also a primary edge.
  \end{itemize}

  When $u_0$ is the source, then $u_1$ is a child of the source which represents a character. Hence $\slink(u_1)$ is the source.
  Thus, although $\slink(u_0)$ is undefined,
  we can use the same argument as above from $u_1$ and from $\slink(u_1)$,
  proving that 
  $(\slink(u_{i-1}), \slink(u_i))$ is a primary edge for every $2 \leq i \leq k-1$.
\end{proof}

Note that Lemma~\ref{lem:LPT_path} does not contain
the edge $(\slink(u_{k-1}), v)$ from the node $\slink(u_{k-1})$
that has the same string label $s_k$
as the last edge $(u_{k-1}, u_{k})$ in the given unary path.
There are two cases for the destination node $v$:
\begin{itemize}
\item[(a)] If the unary path $u_0, u_1, \ldots, u_{k-1}, u_k$ belongs to 
the \emph{shortest} path from the source to $u_k$, then
$v = \slink(u_k)$ and the edge $(\slink(u_{k-1}), \slink(u_k))$ is primary.

\item[(b)] Otherwise,
then $v = u_k$ and the edge $(\slink(u_{k-1}), u_k)$ is secondary.
\end{itemize}
In either case, the sequence of nodes $\slink(u_{0}), \slink(u_1), \ldots, \slink(u_{k-1}), v$ is a path in $\LPTplus(T)$.

We now define the modified fast links on $\simLCDAWG(T)$
in a similar way to Section~\ref{sec:lcdawg}:
Let $(u, a+, v)$ be any $\Plus$-edge of $\simLCDAWG(T)$,
and let $x$ be the string label of this edge.
Then, $\fastlink'(u, a+, v) = \langle u'_1, \ldots, u'_k, v' \rangle$ where
\begin{itemize}
  \item the path $\langle u'_1, \ldots, u'_k, v' \rangle$ spells out $x$, and
  \item $\str(u'_1)$ is the longest proper suffix of $\str(u)$
such that the path $\langle u'_1, \ldots, u'_k, v' \rangle$
contains at least three nodes including $u'_1$ and $v'$ (namely $k \geq 2$).
\end{itemize}
By the suffix property, $\str(v')$ is a suffix of $\str(v)$.

\begin{example}
  See the lower diagrams of Figure~\ref{fig:lcdawg_lpt_plus}.
  \begin{itemize}
  \item For $\Plus$-edge $(u, \mathtt{a}+, z) = (U, Z)$, $\fastlink'(U, Z) = \langle R, A \rangle$ that consists of edges $(R, S)$ and $(S, A)$ on $\LPTplus(T)$.
  \item For $\Plus$-edge $(s, \mathtt{a}+, z) = (S, A)$, $\fastlink'(S, A) = \langle R, B \rangle$ that consists of edges $(R, S)$, $(S, U)$, and $(U, B)$ on $\LPTplus(T)$.
  \item For $\Plus$-edge $(u, \mathtt{c}+, z) = (U, B)$, $\fastlink'(U, B) = \langle R, C \rangle$ that consists of edges $(R, V)$ and $(V, C)$ on $\LPTplus(T)$.
  \end{itemize}
\end{example}

\subsection{Label Extraction and Pattern Matching with $\simLCDAWG(T)$}

This subsection argues how Lemma~\ref{lem:LPT_path} can be used for
edge label extraction and pattern matching on $\simLCDAWG(T)$.

Since the destinations of the fast links are always paths of the tree $\LPTplus(T)$, one can use the same method for string label extraction as $\simLST(T)$ from Section~\ref{sec:sim_LST}, as follows:

\begin{lemma} \label{lem:sim_LCDAWG_edge_label_retrival}
  Given an edge $(u,v)$ of $\simLCDAWG(T)$,
  the string label $x$ of the edge $(u,v)$ can be retrieved in $O(\ell)$ time by using modified fast links, where $\ell = |x|$.
\end{lemma}

\begin{proof}
  By Lemma~\ref{lem:LPT_path},
  extracting an edge label on $\simLCDAWG(T)$
  is reducible to extracting an edge label on $\LPTplus(T)$.
  Since $\LPTplus(T)$ is a tree,
  by applying the same bottom-up method
  to $\LPTplus(T)$ as in the case of Lemma~\ref{lem:sim_LST_edge_label_retrival} for LSTries,
  we can extract the string label of a given edge in $\simLCDAWG(T)$ in linear time in its length.
\end{proof}

Unlike $\LCDAWG(T)$ of Takagi et al.~\cite{Takagi2017},
our $\simLCDAWG(T)$ does not explicitly create a grammar and thus,
naturally, does not use Gasieniec et al.'s data structure~\cite{GasieniecKPS05} for compressed sequential access.

Instead, our $\simLCDAWG(T)$ can use the same pattern matching algorithm as our $\simLST(T)$ from Section~\ref{sec:sim_LST}, as follows:

\begin{lemma} \label{lem:sim_LCDAWG_pattern_matching}
  After $O(\righte(T))$-time preprocessing,
  given a pattern $P$ of length $m$,
  one can find all $\occ$ occurrences of $P$ in the text $T$ in $O(m \log \sigma + \occ)$ time using $\simLCDAWG(T)$.
\end{lemma}

\begin{proof}
  Again, by Lemma~\ref{lem:LPT_path},
  top-down edge label extraction for pattern matching on $\simLCDAWG(T)$
  can be done with $\LPTplus(T)$.
  Since $\LPTplus(T)$ is a tree,
  by applying the same ancestor-based method
  to $\LPTplus(T)$ as in Lemma~\ref{lem:sim_LST_pattern_matching} for $\simLST(T)$,
  we can perform pattern matching in $O(m \log \sigma + \occ)$ time.

  The preprocessing for ancestor queries on $\LPTplus(T)$ takes $O(\righte(T))$ time.
\end{proof}
Refer to Figure~\ref{fig:false_positive} in Section~\ref{sec:sim_LST} for intuition to the proof of Lemma~\ref{lem:sim_LCDAWG_pattern_matching}.
According to the above discussions, if we regard the diagram as a part of $\simLCDAWG(T)$, then the path from $r$ to the node $v'$ is a part of $\LPTplus(T)$.

\begin{figure}[t!]
  \begin{center}
    \includegraphics[scale=0.4]{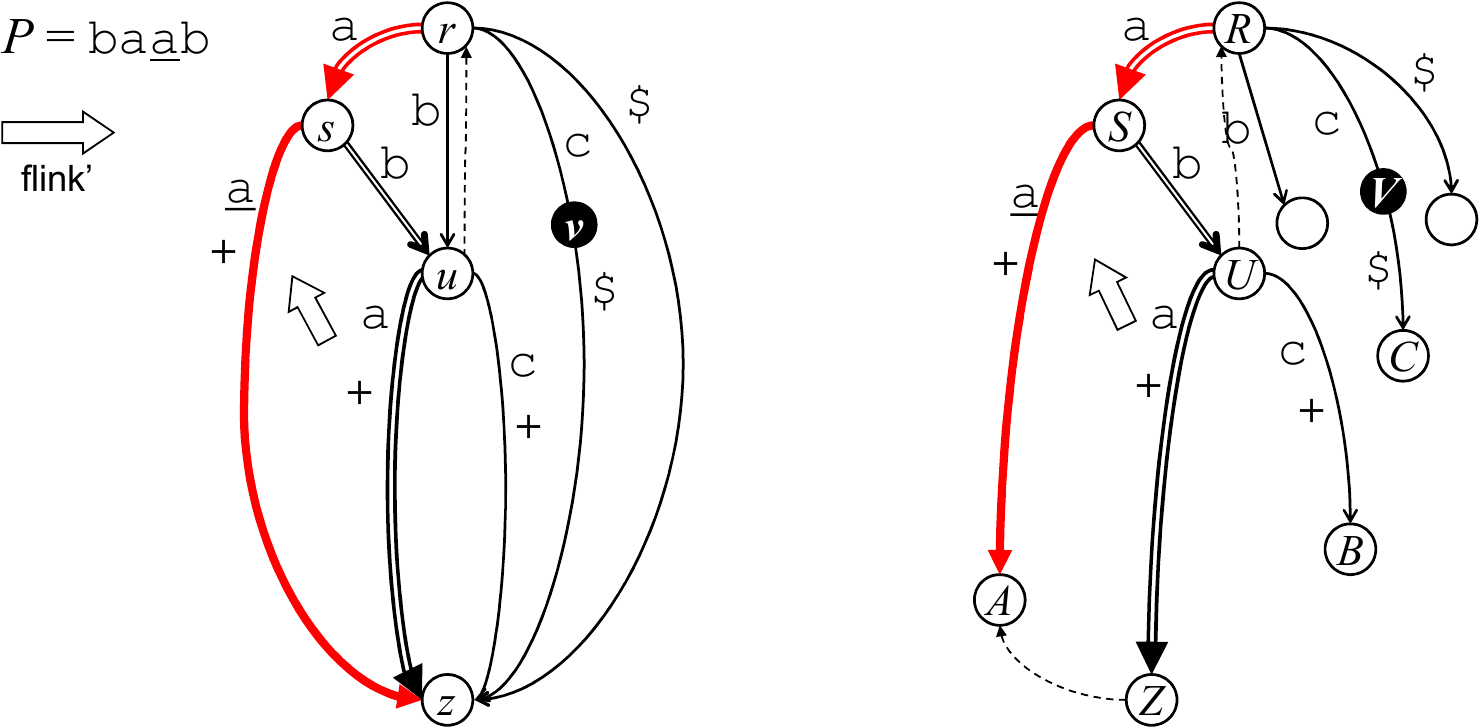}
  \end{center}
  \begin{center}
    \includegraphics[scale=0.4]{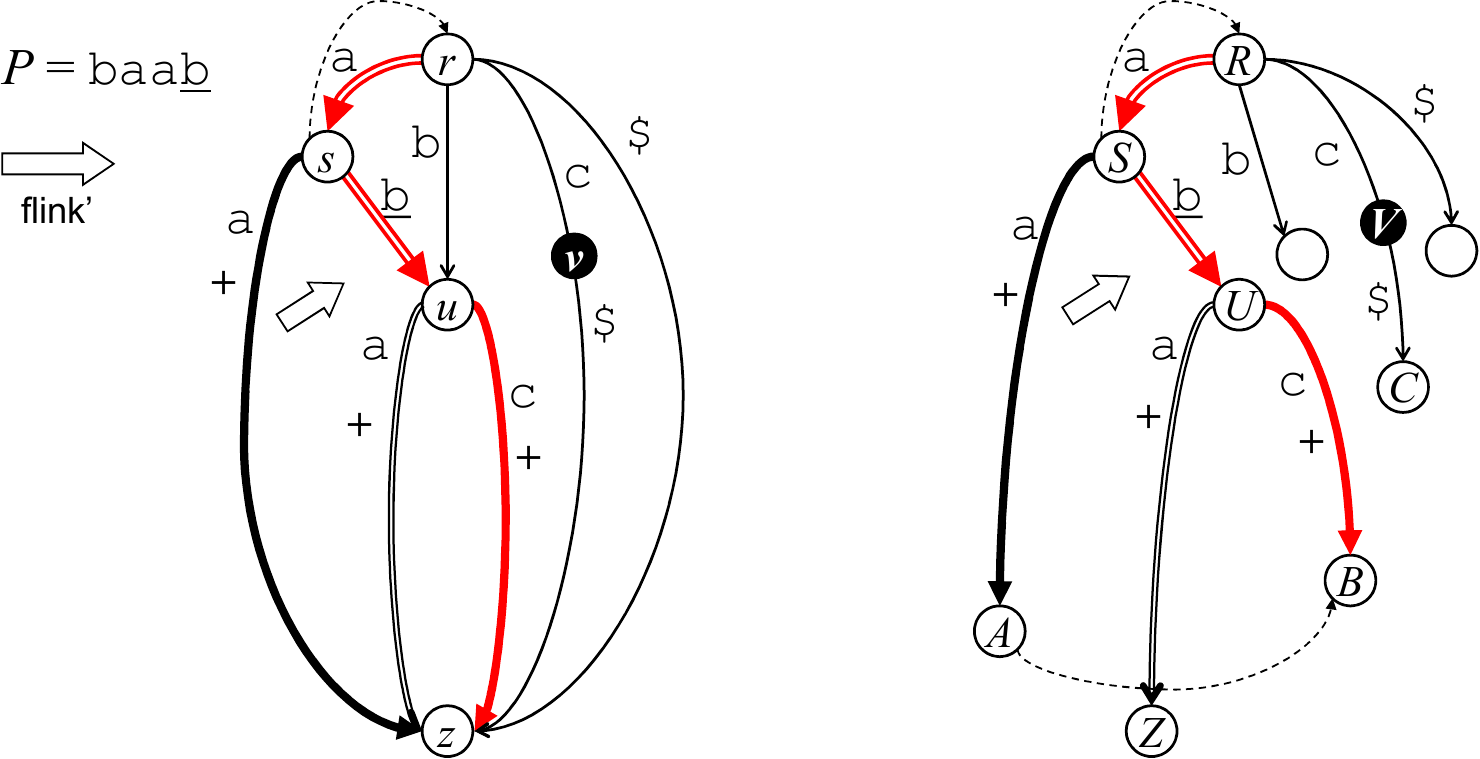}
  \end{center}
  \caption{Matching pattern $P = \mathtt{baab}$ on $\simLCDAWG(T)$ with $T = \mathtt{abaabc\texttt{\$}}$. After traversing $\mathtt{ba}$ from the source, we encounter $+$ (the upper-left diagram). The following character $\underline{\mathtt{a}}$ is decoded by the modified fast link from the edge $(U, Z)$ to the red path $\langle R, A \rangle$ on $\LPTplus(T)$ (the upper-right diagram). We again encounter $+$ on edge $(s, z)$ (the lower-left diagram). The following character $\underline{\mathtt{b}}$ is decoded by the modified fast link from the edge $(S, A)$ to the red path $\langle R, B \rangle$ on $\LPTplus(T)$ (the lower-right diagram).}
  \label{fig:matching_sim_lcdawg}
\end{figure}

\begin{figure}[H]
  \centering
  \includegraphics[scale=0.4]{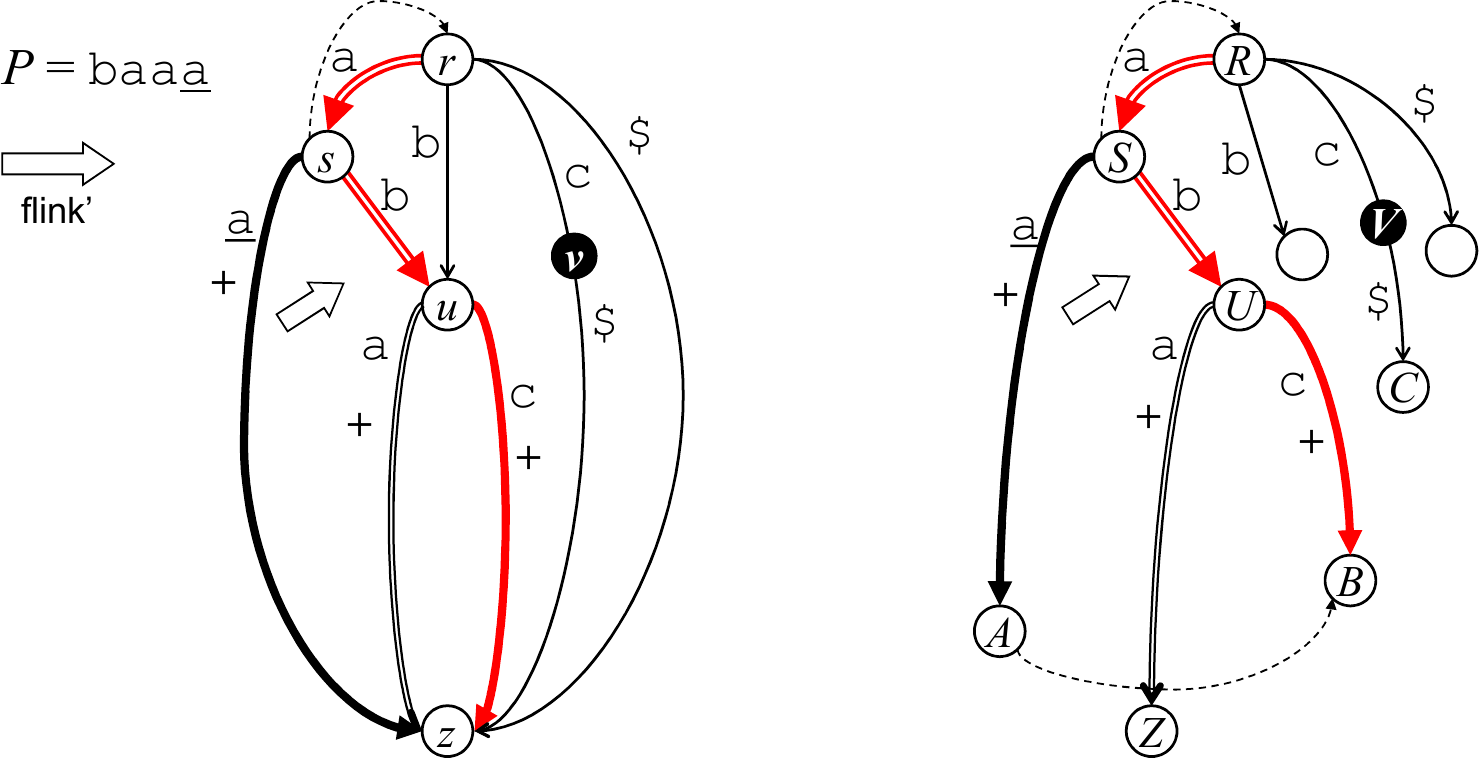}
  \caption{Matching pattern $P = \mathtt{baaa}$ on $\simLCDAWG(T)$ with $T = \mathtt{abaabc\texttt{\$}}$. It is the same as in Figure~\ref{fig:matching_sim_lcdawg} up to $P[1..3] = \mathtt{baa}$. We encounter $+$ on edge $(s, z)$ (the left diagram). The following character $\underline{\mathtt{a}}$ is detected from node $S$ on $\LPTplus(T)$. However, since it is not on the red path, $P = \mathtt{baaa}$ does not occur in $T$.}
  \label{fig:failed_matching_sim_lcdawg}
\end{figure}

Figure~\ref{fig:matching_sim_lcdawg} and Figure~\ref{fig:failed_matching_sim_lcdawg} illustrate examples of pattern matching using $\simLCDAWG(T)$, where the pattern $P$ occurs (resp. $P$ does not occur) in $T$.

\subsection{Construction of $\simLCDAWG(T)$}

Below we show how to build $\simLCDAWG(T)$ with modified fast links,
from the edge-sorted $\CDAWG(T)$ augmented with suffix links
together with the input string $T$.
Our goal is to work in space $O(\lefte(T))$ linear in the size of $\CDAWG(T)$
(except for the string $T$).
The challenging part of this construction is that
we are not allowed to build $\LCDAWG(T)$ since
it can require $\Omega(\righte(T) \sqrt{n})$ space
due to Lemma~\ref{lem:lowerbound_el_er}.
Still, we can build our $\simLCDAWG(T)$ efficiently as follows:

\begin{lemma} \label{lem:simLCDAWG_construction_from_CDAWG}
  Given the edge-sorted $\CDAWG(T)$ of size $\righte(T)$ with suffix links
  and the input string $T$,
  one can build $\simLCDAWG(T)$ with modified fast links
  in $O(\righte(T) \log \sigma)$ time
  and $O(\heightSLT(T))$ working space,
  where $\heightSLT(T)$ denotes the height of the suffix link tree of $\CDAWG(T)$.
\end{lemma}

\begin{proof}
  We first insert the children of the source trivially in $O(\sigma) \subseteq O(\righte(T))$ time. This gives the nodes and edges of $\simLCDAWG(T)$.
  
  We can compute the modified fast links as follows.
  Let $(u, (a, (i,j)), v)$ be an arbitrarily chosen edge of length $j-i+1 > 1$
  of which the fast link has not been computed yet.
  Let $x = T[i..j]$ be the string label of this edge.
  We traverse the two suffix link chains from $u$ and from $v$ in parallel
  and take the edges beginning with $a$,
  until we find the first pair $\slink^k(u), \slink^h(v)$ of nodes
  (i.e. with the smallest $k, h$ such that $0 \leq h \leq k$)
  in the two suffix link chains that satisfy one of the followings (see also Figure~\ref{fig:simLCDAWG_construction} for illustration):
  \begin{enumerate}
  \item[(1)] $\slink^k(u)$ is a non-parent ancestor of $\slink^h(v)$ in $\LPTplus(T)$, or
  \item[(2)] $(\slink^k(u), \slink^h(v))$ is an edge of $\simLCDAWG(T)$ for which
  the modified fast link $\fastlink'(\slink^k(u), \slink^h(v))$ has already been computed.
  \end{enumerate}  

\begin{figure}[h]
	\centering
        \includegraphics[scale=0.5]{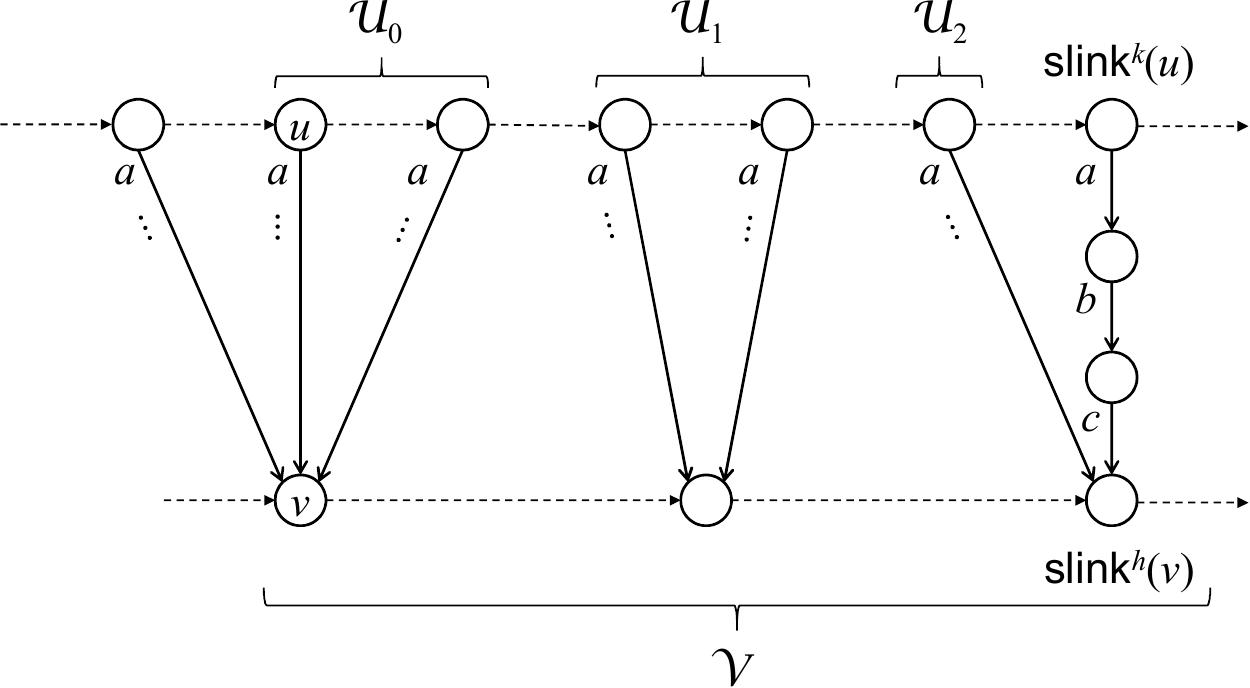}
	\caption{Illustration for the proof of Lemma~\ref{lem:simLCDAWG_construction_from_CDAWG}, where $k = 5$ and $h = 2$ in this diagram. The modified fast links of the edges found in the parallel suffix link chains from $u$ to $\slink^k(u)$ and from $v$ to $\slink^h(u)$ are set to point to the path of $\LPTplus(T)$ that spells out $x$ from $\slink^k(u)$, where $x$ is the string label of the edge $(u, v)$.}
	\label{fig:simLCDAWG_construction}
\end{figure}
  
  Let us first consider Case (1).
  Let
  \begin{eqnarray*}
    \mathcal{U} & = & \{u = \slink^0(u), \slink^1(u), \ldots, \slink^{k-1}(u)\}, \\
    \mathcal{V} & = & \{v = \slink^0(v), \slink^1(v), \ldots, \slink^h(v)\}.
  \end{eqnarray*}
  We partition $\mathcal{U}$ into $h+1$ disjoint subsets $\mathcal{U}_0, \ldots, \mathcal{U}_h$,
  such that for any node $x \in \mathcal{U}_i$~($0 \leq i \leq h$),
  $(x, \slink^i(v))$ is an edge of $\simLCDAWG(T)$.
  Then, we create the fast links from the edges $(x, \slink^i(v))$
  for all nodes $x$ in $\mathcal{U}_i$ to
  the path of $\LPTplus(T)$ from $\slink^k(u)$ that spells out the string label $x$.
  We find this path from $\slink^k(u)$ by the so-called skip-and-count method
  (c.f.~\cite{Ukkonen1995,Crochemore2016})
  toward $\slink^h(v)$ on $\simLCDAWG(T)$, using $O(\log \sigma)$ time for each edge in the path.
  

  In Case (2),
  the two sets $\mathcal{U}$ and $\mathcal{V}$ can be defined with a slight modification such that
  $(\slink^k(u), \slink^h(v))$ is the first pair of nodes in the suffix link chains, of which the modified fast link $\fastlink'((\slink^k(u), \slink^h(v)))$ has already been computed.
  Then, we set the modified fast links of all edges $(x, \slink^i(v))$
  with $x \in \mathcal{U}_i$ so that it points to the destination node of $\fastlink'((\slink^k(u), \slink^h(v)))$ for all $0 \leq i \leq h$.

  Both in Cases (1) and (2),
  at least one new fast link is computed in each suffix link traversal.
  Thus, the modified fast link of each edge can be computed in $O(\log \sigma)$ amortized time.
  Overall, we can compute the modified fast links of all edges in a total of $O(\righte(T) \log \sigma)$ time.

  The working space for computing the modified fast links is linear in the maximum length of the suffix link chains, which is bounded by $\heightSLT(T)$.
  We remark that $\heightSLT(T) \leq |\VCDAWG|$ always holds.

Lastly, we modify each edge label $(a, (i,j))$ to $a$ if $j-i+1 = 1$, or to $a+$ if $j-i+1 > 1$, in $O(\righte(T))$ total time with $O(1)$ working space.
\end{proof}

When our input is the string $T$ only, we have the following constructions:

\begin{corollary} \label{coro:simLCDAWG_construction_from_string}
  Given a string $T$ of length $n$,
  one can build $\simLCDAWG(T)$ 
  \begin{enumerate}
  \item[(A)] in $O(n \log \sigma)$ time with $O(\righte(T))$ working space for general ordered alphabets, or
  \item[(B)] in $O(n)$ time with $O(n)$ working space for integer alphabets of polynomial size in $n$.
  \end{enumerate}
\end{corollary}

\begin{proof}
  It is known that the edge-sorted $\CDAWG(T)$ with suffix links can be built
  in $O(n \log \sigma)$ time with $O(\righte(T))$ working space~\cite{InenagaHSTAMP05}. Combining this with Lemma~\ref{lem:simLCDAWG_construction_from_CDAWG}, we readily obtain Construction (A).

  For Construction (B), we first build $\CDAWG(T)$ in $O(n)$ time and space using the algorithm of Narisawa et al.~\cite{NarisawaHIBT17}, which first constructs $\STree(T)$ and then merges its isomorphic subtrees by following suffix links. The crux is that the incoming edges of each node of the resulting $\CDAWG(T)$ are sorted in the order of the length of the paths from the source. This permits us to remove the $\log \sigma$ factor in the time complexity that is required to search for the edges beginning with the same characters. This is because we can know which node of $\LPTplus(T)$ corresponds to the last edge of the path from $\slink^k(u)$ to $\slink^h(v)$ spelling out the substring $x = T[i..j]$.
  This gives us Construction (B).
\end{proof}

%% file: conclusions.tex
\section{Conclusions and Open Problems}
\label{sec:conclusions}

This paper proposed a new string indexing structure
called the \emph{simplified linear-size suffix trie} $\simLST(T)$
that is a simpler alternative to the existing linear-size suffix trie $\LST(T)$.
The main advantage of our new $\simLST(T)$ is that
it does not require type-2 non-branching nodes which are the key augmentations of $\LST(T)$,
except those that are the children of the root.
We showed that standard preprocessing on the tree
for constant-time ancestor queries between two nodes can replace
the type-2 nodes.
We then provided matching upper and lower bounds on the number of nodes in $\simLST(T)$,
and presented efficient algorithms for edge label extraction, pattern matching, and construction.

This idea was then extended to our second new string indexing structure
called the \emph{simplified LCDAWG} $\simLCDAWG(T)$ of size $O(\righte(T))$ that is a more space-efficient alternative to the existing $\LCDAWG(T)$ of size $O(\lefte(T) + \righte(T))$.
The key idea was the use of $\LPTplus(T)$ that is an extended tree of the longest-path spanning tree $\LPT(T)$ of $\CDAWG(T)$.

An intriguing open question is whether one can remove the $\log \sigma$ factor
in the construction time of $\simLCDAWG(T)$ of Lemma~\ref{lem:simLCDAWG_construction_from_CDAWG}, and without the input string $T$.
If we can assume that the in-coming edges of each node are sorted in decreasing order of the lengths of the paths from the source, then we are able to achieve the above.
Can we do this without such an assumption?

Another interesting open question is whether one can construct simLSTries and/or simLCDAWGs in an online manner.
We believe that by using the dynamic LCA data structure~\cite{ColeH05} in place of the standard preprocessing for ancestor queries, one can build simLSTries online by a modification to Ukkonen's algorithm~\cite{Ukkonen1995} for suffix tree online constructions,
and can build simLCDAWGs online by a modification to Inenaga et al.'s algorithm~\cite{InenagaHSTAMP05} for CDAWG online constructions.
However, since our simLSTries and simLCDAWGs themselves work on the pointer machine model,
it would be interesting to do this without the dynamic LCA structure, that works only on the word RAM model.